      \documentclass[twocolumn,showpacs,preprintnumbers,amsmath,amssymb]{revtex4-2}
   \usepackage{graphicx,xcolor}    
   \usepackage{epstopdf,epsfig}
   \usepackage{bm}
   \usepackage{dcolumn}
   \usepackage{natbib}
   \usepackage{physics}

\begin{document}
   	
   	\title{Excitonic effects in twisted bilayer graphene}
   	
   	\author{V. Apinyan\footnote{Corresponding author. Tel.:  +48 71 3954 284; E-mail address: v.apinyan@int.pan.wroc.pl.}, T. K. Kope\'{c}}
   	\affiliation{Institute of Low Temperature and Structure Research, Polish Academy of Sciences\\
   		PO. Box 1410, 50-950 Wroc\l{}aw 2, Poland \\}
   	
   	\date{\today}

   	\begin{abstract}
%
In the present work, we consider the excitonic effects in the twisted bilayer graphene (tBLG) within the rotated bilayer Hubbard model. Both, intralayer and interlayer Coulomb interactions have been considered and the half-filling condition is imposed for the electronic densities is both layers of the bilayer. We calculate the excitonic pairing gap parameter and the chemical potential for different twist angles and different values of the interlayer Coulomb interaction parameter. Furthermore, we show the appearance of the electronic flat bands in the electronic band structure, mediated by the excitonic effects. We show that there is a doubling effect of the Dirac's $K$-point at the low interaction limit and one of Dirac's nodes is stable and the other one changes its position as a function of rotation angle. At the large twist angle limit, there appear two additional Dirac-like nodes at the $M$-point in the Brillouin zone. We show the excitonic red-shift effect of the principal Dirac's point $K$, in the low interaction limit, while, at the strong interactions, we get also the blue-shift effect at the $M$-point. Apart from the mentioned effects, the theory evaluated here predicts a metal-semiconductor transition in the tBLG system when augmenting the interlayer Coulomb interaction parameter. 
   	\end{abstract}

   	\maketitle

   	\renewcommand\thesection{\arabic{section}}
   	
   	\section{\label{sec:Section_1} Introduction}
   	%
   	The tremendous attention has been put recently on the physical effects in the twisted bilayer graphene (tBLG) due to its spectacular properties undiscovered in other bilayer structures and two-dimensional materials. Particularly, inducing the twist between the layers in the bilayer graphene leads to the formation of the moir\'e-like superlattice structures \cite{cite_1, cite_2, cite_3, cite_4, cite_5, cite_6, cite_7, cite_8, cite_9}. It has been shown both experimentally \cite{cite_10, cite_11, cite_12, cite_13} and theoretically \cite{cite_14, cite_15, cite_16, cite_17, cite_18, cite_19, cite_20, cite_21} that when this angle is close to the so-called magic angles, the electronic band structure becomes flat near the zero Fermi energy, which is a direct consequence of the strong interlayer correlations in the twisted bilayer graphene (tBLG). The flat bands lead to the Mott-like insulating states with the intervalley coherence, at the half-filling, arising at the magic twist angles \cite{cite_16, cite_17}. 
       The excitonic effects in the layered two-dimensional structures \cite{cite_22} represent another important interest due to their possible direct applications in various fields of the modern material nanosciences, including photovoltaics and photocatalytic \cite{cite_23}, technological use as nanometre-scale light sources \cite{cite_24}, exciton-valleytronics \cite{cite_25} and photodetectors. Recently a low-energy effective model has been used \cite{cite_26} to found the formation of bounded excitonic states with the significant binding energy of the order of $0.5$ eV in tBLG, which is an order of magnitude larger than that found in metallic carbon nanotubes and which is explained by the formation of ghost Fano resonance peaks \cite{cite_27}.
       Bright and dark excitonic states and the peak-splitting are observed in tBLG \cite{cite_28} using two-photon photoluminescence (PL) and intraband transient absorption spectra. The associated stable interlayer exciton binding
       energy ranges from $0.5$ to $0.7$ eV for the twist angles $\theta = 8^\circ$ to $17^\circ$. Actually, there exist two principal mechanisms that explain the mentioned resonant excitonic absorption in tBLG; the hot electron van Hove singularity (vHs) model, and the formation of bound interlayer excitons \cite{cite_29, cite_30, cite_31, cite_32, cite_33, cite_34, cite_35}. Particularly, the foremost of them defines two degenerate optical transitions occurring between band flattened anti-crossing regions in the scheme of the interlayer band structure overlap.
       Recently, a series of interesting many-body electronic physics occur in anticrossing regions of the tBLG band structure, such as exciton effects, $\theta$-dependent superconductivity and metal-insulator
       transitions \cite{cite_36, cite_37}. Recently, it has been shown theoretically \cite{cite_38} that an applied bias voltage leads to the appearance of two hole-like and two electron-like Fermi surface sheets in tBLG with a perfect nesting among those components and leads to the formation of the excitonic gap in the band structure picture. 
   	It has been shown in Ref.\onlinecite{cite_38} that the gap depends on the twist angle and can be varied by changing the bias voltage and leading to the spin-density-wave order in tBLG. Those results appear to be consistent with the recent experiments given in Ref.\onlinecite{cite_39}. The sharpness of the two-dimensional saddle-point excitons is observed in tBLG \cite{cite_40} by optical reflectivity and Raman scattering methods leading to the resonances with Stokes and anti-Stokes Raman emission components with the dependence on the excitation power and twist angle $\theta$.
       
       In the present paper, we show that the excitonic pair formations with the positive value of the spin-symmetric gap parameter occur only in a certain interval of twisting angle between the layers of the bilayer graphene. We discuss the excitonic pair formation in the strongly interacting twisted bilayer graphene structure within the generalized bilayer Hubbard model for tBLG. Both intralayer and interlayer Coulomb interaction parameters have been included in the calculations and the excitonic gap parameter is calculated self-consistently as a function of the interlayer Coulomb interaction parameter. We show the non-dependence of the gap on the intralayer Coulomb interaction in good accordance with the previous works on the subject of the excitonic effects in AB-stacked bilayer graphene \cite{cite_41, cite_42}. Furthermore, the calculations of the electronic band structure in tBLG show the existence of the flat bands formed from the inner energy levels in the band structure. We present the calculations of the electronic band structure in tBLG, both at the low and strong interlayer Coulomb interaction limit and for small and high twist angles between the layers. The red-shift effect of the Dirac's $K$-point is discussed in the paper and the appearance of doubled and quartered Dirac's nodes on the ${\bf{k}}$-axes is shown in the low interaction limit. Among the obtained results, we show the existence of the metal-semiconductor transition in the tBLG at the strong values of the interlayer interaction parameter $W$ and independently of the twist angle $\theta$.     
   	%
   	\section{\label{sec:Section_2} The Hubbard model for twisted bilayer graphene}
   	%
   	\begin{figure}
   		\begin{center}
   			\includegraphics[scale=0.9]{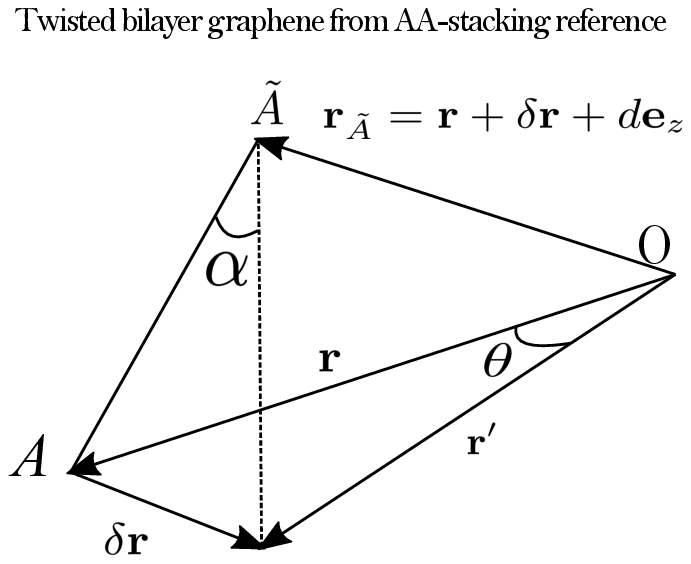}
   			\caption{\label{fig:Fig_1}(Color online) The schematics of twisted bilayer graphene structure. The AA-stacked bilayer graphene is considered as the reference, i.e., the case $\theta=0$.}
   		\end{center}
   	\end{figure} 
   	%
   	For studying the excitonic effects in twisted bilayer graphene system we consider the bilayer generalization of the Hubbard model with the half-filling condition for the electronic densities in each layer of the BLG. Initially, we suppose the AA-stacked configuration for the bilayer graphene. Let's attribute the notations $a$ and $b$ for the electron destruction operators at different non-equivalent sublattice sites $A$ and $B$ on the honeycomb lattice in the bottom layer 1, while the notations $\tilde{a}$ and $\tilde{b}$ indicate the electron destruction operators on the lattice sites $\tilde{A}$ and $\tilde{B}$, in the top layer 2. The notations ${a}^{\dag}$, ${b}^{\dag}$, ${\tilde{a}}^{\dag}$ and ${\tilde{b}}^{\dag}$ correspond to the electron creation operators on the indicated lattice site positions. Then the intralayer Hamiltonian of the tBLG system will be written as
   	\begin{eqnarray}
   	H_{||}=&&-\gamma_{0}\sum_{\left\langle {\bf{r}}{\bf{r}}'\right\rangle,\sigma}\left[{a}^{\dag}_{\sigma}({\bf{r}})b_{\sigma}({\bf{r}}')+h.c.\right]
   	\nonumber\\
   	&&-\gamma_{0}\sum_{\left\langle {\bf{r}}{\bf{r}}'\right\rangle,\sigma}\left[{\tilde{a}}^{\dag}_{\sigma}({\bf{r}})\tilde{b}_{\sigma}({\bf{r}}')+h.c.\right]
   	\nonumber\\
   	&&-\sum_{{\bf{r}}\sigma}\sum_{\ell=1,2}\mu_{\ell}n_{\ell\sigma}({\bf{r}}).
   	\label{Equation_1}
   	\end{eqnarray}
   	Here, we have added the chemical potential term in the Hamiltonian (see the last tern in Eq.(\ref{Equation_1})). 
   	The parameter $\gamma_0$ in Eq.(\ref{Equation_1}) is the intralayer hopping amplitude which is assumed the same for both layers in the BLG. The summation $\left\langle {\bf{r}}{\bf{r}}'\right\rangle$ runs over the nearest neighbour lattice sites, $\sigma$ is the spin index which takes two values $\sigma=\uparrow, \downarrow$. The index $\ell$, in the last term in Eq.(\ref{Equation_1}) denotes the layers in the BLG. Next, $\mu_{\ell}$ is the chemical potential in the layer $\ell$ and $n_{\ell\sigma}$ is the total electron density operator in the given layer in the BLG, i.e., $n_{\ell\sigma}({\bf{r}})={a}^{\dag}_{\sigma}({\bf{r}}){a}_{\sigma}({\bf{r}})+{b}^{\dag}_{\sigma}({\bf{r}}){b}_{\sigma}({\bf{r}})$ for $\ell=1$ and $n_{\ell\sigma}({\bf{r}})={\tilde{a}}^{\dag}_{\sigma}({\bf{r}})\tilde{a}_{\sigma}({\bf{r}})+{\tilde{b}}^{\dag}_{\sigma}({\bf{r}})\tilde{b}_{\sigma}({\bf{r}})$ for $\ell=2$.
   	The Hamiltonian in Eq.(\ref{Equation_1}) could be rewritten in a more convenient form. We have
   	\begin{eqnarray}
   	H_{||}=&&-\gamma_{0}\sum_{ {\bf{r}},\sigma}\sum^{3}_{i=1}\left[{a}^{\dag}_{\sigma}({\bf{r}})b_{\sigma}({\bf{r}}+\bm{\mathit{\delta}}_{i})+h.c.\right]
   	\nonumber\\
   	&&-\gamma_{0}\sum_{{\bf{r}},\sigma}\sum^{3}_{i=1}\left[{\tilde{a}}^{\dag}_{\sigma}({\bf{r}})\tilde{b}_{\sigma}({\bf{r}}+\bm{\mathit{\delta}}'_{i})+h.c.\right]
   	\nonumber\\
   	&&-\sum_{{\bf{r}}\sigma}\sum_{\ell=1,2}\mu_{\ell}n_{\ell\sigma}({\bf{r}}),
   	\label{Equation_2}
   	\end{eqnarray}
   	where the vectors $\bm{\mathit{\delta}}_{i}$ and $\bm{\mathit{\delta}}'_{i}$ with $i=1,..3$, in the first and second terms in Eq.(\ref{Equation_2}) are the nearest neighbour vectors in the layer 1 and layer 2, respectively. For the layer 1, we have $\bm{\mathit{\delta}}=\left(\bm{\mathit{\delta}}_1,\bm{\mathit{\delta}}_2,\bm{\mathit{\delta}}_3\right)$, where $\bm{\mathit{\delta}}_1=\left(a/2,a\sqrt{3}/2\right)$, $\bm{\mathit{\delta}}_2=\left(a/2,-a\sqrt{3}/2\right)$ and $\bm{\mathit{\delta}}_3=\left(-a,0\right)$. We suppose that the layer 2 is rotated counterclockwise through an angle $\theta$ about the layer 1. As a usual rotation in the Euclidean space we can write 
   	\begin{eqnarray}
   	\bm{\mathit{\delta}}'_i={\cal{R}}(\theta)\bm{\mathit{\delta}}_i,
   	\label{Equation_3}
   	\end{eqnarray}
   	where ${\cal{R}}(\theta)$ is the 2D rotation matrix, which rotates each vector $\bm{\mathit{\delta}}_i$ through an angle $\theta$. Thus, we have for the nearest neighbour vectors in the layer 2 $\bm{\mathit{\delta}}'_1=\left(a\cos(\theta)/2-a\sqrt{3}\sin(\theta)/2, a\sin(\theta)/2+a\sqrt{3}\cos(\theta)/2\right)$, $\bm{\mathit{\delta}}'_2=\left(a\cos(\theta)/2+a\sqrt{3}\sin(\theta)/2, a\sin(\theta)/2-a\sqrt{3}\cos(\theta)/2\right)$ and $\bm{\mathit{\delta}}'_3=\left(-a\cos(\theta),-a\sin(\theta)\right)$.  
   	
   	The interlayer hopping term is given by the following Hamiltonian
   	\begin{eqnarray}
   	H_{\perp}=&&-\tilde{\gamma}_1(|\delta{{\bf{r}}}|)\sum_{{\bf{r}},\sigma}\left[a^{\dag}_{\sigma}({\bf{r}})\tilde{a}({\bf{r}}+\delta{{\bf{r}}}+d{\bf{e}}_{z})+h.c.\right]
   	\nonumber\\
   	&&-\tilde{\gamma}_1(|\delta{{\bf{r}}}|)\sum_{{\bf{r}},\sigma}\left[b^{\dag}_{\sigma}({\bf{r}})\tilde{b}({\bf{r}}+\delta{{\bf{r}}}+d{\bf{e}}_{z})+h.c.\right].
   	\nonumber\\
   	\label{Equation_4}
   	\end{eqnarray}
   	The twisted vector ${\bf{r}}_{\tilde{A}}={\bf{r}}+\delta{\bf{r}}+d{\bf{e}}_{z}$ is presented in Fig.~\ref{fig:Fig_1}, where $d$ is the interlayer separation $d=3.35$ $\mathrm{\AA}$.  The interlayer hopping parameter is given by the function $\gamma_1(|\delta{{\bf{r}}}|)$, where $|\delta{{\bf{r}}}|$ is constant for a given angle $\theta$, i.e., $|\delta{{\bf{r}}}|=2\sin(\theta/2)a$, where $a$ is the lattice constant in graphene: $a=1.42$ $\mathrm{\AA}$.   
   	The hopping amplitude $\gamma_1(|\delta{{\bf{r}}}|)$ can be expressed via the Slater-Koster parametrization and using the functions potentials $V_{pp\sigma}$ and $V_{pp\pi}$ \cite{cite_1}
   	\begin{eqnarray}
   	\tilde{\gamma}_1(|\delta{{\bf{r}}}|)=\cos^{2}(\alpha)V_{pp\sigma}(\sqrt{d^{2}+|\delta{{\bf{r}}}|^{2}})
   	\nonumber\\
   	+\sin^{2}(\alpha)V_{pp\pi}(\sqrt{d^{2}+|\delta{{\bf{r}}}|^{2}}).
   	\label{Equation_5}
   	\end{eqnarray}
   	Here, the functions $V_{pp\sigma}$ and $V_{pp\pi}$ depend only on the distance between two sites and $\alpha$ is the angle between the $z$ axis and line $A\tilde{A}$ connecting the sites in the positions ${\bf{r}}$ and ${\bf{r}}'+d{\bf{e}}_{z}$. We will use the exponentially decreasing functions $V_{pp\sigma}$ and $V_{pp\pi}$, considered in Refs.\onlinecite{cite_2, cite_3}, which are given as
   	\begin{eqnarray}
   	V_{pp\sigma}(r)=\gamma_1F_{c}(r)\exp\left[q_{\sigma}(1-\frac{r}{d})\right],
   	\nonumber\\
   	V_{pp\pi}(r)=\gamma_0F_{c}(r)\exp\left[q_{\pi}(1-\frac{r}{a})\right],
   	\label{Equation_6}
   	\end{eqnarray}
   	where $\gamma_1$ is the interlayer hopping amplitude of the untwisted BLG, and also we have for the spatial exponential-decreasing coefficients the following relation ${q_{\pi}}/{a}={q_{\sigma}}/{d}$.
   	The function $F_{c}(r)$ in Eq.\ref{Equation_6} is so-called cut-off function which reduces the long-range hopping amplitudes \cite{cite_3} and we have
   	\begin{eqnarray}
   	F_{c}(r)=\frac{1}{1+\exp\left[\frac{\left(r-r_c\right)}{l_c}\right]},
   	\label{Equation_7}
   	\end{eqnarray}
   	where $r_c=2.5\sqrt{3}a=6.14 $ $\mathrm{\AA}$  and $l_c=0.265$ $\mathrm{\AA}$ \cite{cite_3}. 
   	Furthermore, we include the intralayer and interlayer Coulomb interaction terms. Namely, we have for the interaction part of the system the following Hubbard Hamiltonian
   	\begin{eqnarray}
   	&&H_{U-W}=U\sum_{{\bf{r}}}\sum_{\zeta}\left[\left(n_{\zeta\uparrow}-1/2\right)\left(n_{\zeta\downarrow}-1/2\right)-1/4\right]
   	\nonumber\\
   	&&+W_{\theta}\sum_{{\bf{r}}\sigma\sigma'}\sum_{\eta,\tilde{\eta}}\kappa_{\eta\tilde{\eta}}\left[\left(n_{\eta\sigma}({\bf{r}})-1/2\right)\left(n_{\tilde{\eta}\sigma'}({\bf{r}}+\delta{\bf{r}}+d{\bf{e}}_{z})\right.\right.
   	\nonumber\\
   	&&\left.\left.-1/2\right)-1/4\right].
   	\nonumber\\
   	\label{Equation_8}
   	\end{eqnarray}
   	The summation indices are $\zeta=a,b,\tilde{a},\tilde{b}$, $\eta=a,b$, $\tilde{\eta}=\tilde{a},\tilde{b}$. The coefficient $\kappa_{\eta\tilde{\eta}}$, in the second term in Eq.(\ref{Equation_8}) is defined in such a way that $\kappa_{\eta\tilde{\eta}}=1$, if the $\eta=a$ and $\tilde{\eta}=\tilde{a}$, or $\eta=b$ and $\tilde{\eta}=\tilde{b}$ simultaneously, otherwise, we have $\kappa_{\eta\tilde{\eta}}=0$. 
       Parameters $U$ and $W_{\theta}$, in Eq.(\ref{Equation_8}), signify local intralayer and twisted interlayer Coulomb interactions in the considered twisted BLG structure. Particularly, the parameter $W_{\theta}$ is defined as the interaction parameter between the electrons on the sites $A$ (or $B$), in the bottom layer, and the adjacent twisted sites $\tilde{A}$ (or $\tilde{B}$), in the upper layer. By considering, initially, $AA$-stacked bilayer graphene configuration, as the reference, we can express the interaction parameter $W_{\theta}$ with the help of the interaction parameter $W$ of the $AA$-stacked reference configuration and the angle $\alpha$ (see in Fig.~\ref{fig:Fig_1}). Namely, we have
   	\begin{eqnarray}
   	W_{\theta}=W\cos{\alpha}=\frac{W}{\sqrt{1+4\left(\frac{a}{d}\right)^{2}\sin^{2}(\theta/2)}}.
   	\label{Equation_9}
   	\end{eqnarray}
   	Next, the total Hamiltonian of the twisted bilayer graphene system is given by 
   	\begin{eqnarray}
   	H_{\rm tBLG}=H_{||}+H_{\perp}+H_{U-W},
   	\label{Equation_10}
   	\end{eqnarray}
   	where all terms $H_{||}$, $H_{\perp}$ and $H_{U-W}$, defined in Eqs.(\ref{Equation_2}), (\ref{Equation_4}) and $\ref{Equation_8}$, are properly included. 
   	%
\section{\label{sec:Section_2} Hubbard-Stratanovich linearization and fermionic action}
%
\subsection{\label{sec:Section_2_1} Partition function and the nonlinear density terms}
%
Here, we will show how the interaction terms will be handled in the fermionic-field path integral formalism \cite{cite_42}. For this, we will use furthermore the Grassmann representation for fermionic variables, and we write the partition function of the system in the imaginary time fermion path integral formalism. We introduce the imaginary-time variables $\tau$ \cite{cite_43}, at each lattice site ${\bf{r}}$. The time variables $\tau$ vary in the interval $(0,\beta)$, where $\beta=1/T$ with $T$ being the thermodynamic temperature. The grand canonical partition function of the system is
\begin{eqnarray}
Z=\int\left[D\bar{X}DX\right]\left[D\bar{Y}DY\right]e^{-S\left[\bar{X},X,\bar{Y},Y\right]},
\label{Equation_11}
\end{eqnarray}
and the fermionic action $S\left[\bar{X},X,\bar{Y},Y\right]$ is given as follows
\begin{eqnarray}
S\left[\bar{X},X,\bar{Y},Y\right]=\sum_{l=1,2}S^{(l)}_{\rm B}\left[\bar{X},X\right]
\nonumber\\
+\sum_{l=1,2}S^{(l)}_{\rm B}\left[\bar{Y},Y\right]+\int^{\beta}_{0}d\tau H_{\rm tBLG}\left(\tau\right).
\label{Equation_12}
\end{eqnarray}
Here, the first two terms are the Berry terms for the layers with the indices $\ell=1,2$
\begin{eqnarray}
S^{(l)}_{\rm B}\left[\bar{X},X\right]=\sum_{{\bf{r}},\sigma}\int^{\beta}_{0}d\tau \bar{X}_{l,\sigma}({\bf{r}},\tau)\frac{\partial}{\partial \tau}X_{l,\sigma}({\bf{r}},\tau),
\label{Equation_13}
\newline\\
S^{(l)}_{\rm B}\left[\bar{Y},Y\right]=\sum_{{\bf{r}},\sigma}\int^{\beta}_{0}d\tau \bar{Y}_{l,\sigma}({\bf{r}},\tau)\frac{\partial}{\partial \tau}Y_{l,\sigma}({\bf{r}},\tau),
\label{Equation_14}
\end{eqnarray}
where we have introduced the following notations for the fermionic operators: $X_{1,\sigma}({\bf{r}},\tau)=a_{\sigma}({\bf{r}},\tau)$, $X_{2,\sigma}({\bf{r}},\tau)=\tilde{a}_{\sigma}({\bf{r}},\tau)$, $Y_{1,\sigma}({\bf{r}},\tau)=b_{\sigma}({\bf{r}},\tau)$ and $Y_{2,\sigma}({\bf{r}},\tau)=\tilde{b}_{\sigma}({\bf{r}},\tau)$. The Hamiltonian $H\left(\tau\right)$, in the last term, in Eq.(\ref{Equation_12}), is the Hamiltonian of the interacting twisted BLG system, given in Eq.(\ref{Equation_10}), above. 

Next, we consider the nonlinear Hubbard interaction terms in Eq.(\ref{Equation_8}) and we linearize them in the partition function by using the Hubbard-Stratanovich decoupling procedure. For the product of electron density operators with the opposite spin directions, in the  non-linear intralayer $U$-term, we can write 
\begin{eqnarray}
n_{\zeta\uparrow}n_{\zeta\downarrow}=\frac{n^{2}_{\zeta}}{4}-S^{2}_{\zeta{z}},
\label{Equation_15}
\end{eqnarray}
where $S_{\zeta{z}}$ is the $z$-component of the generalized spin operator ${\bf{S}}_{\zeta}({\bf{r}},\tau)=1/2\sum_{\alpha,\beta = \uparrow, \downarrow}\bar{\zeta}_{\alpha}({\bf{r}}\tau)\hat{\sigma}_{\alpha\beta}\zeta_{\beta}({\bf{r}},\tau)$, for different sublattices, in the layers of the BLG structure. It is defined as $S_{\zeta}({\bf{r}},\tau)=1/2\left(\zeta_{\uparrow}({\bf{r}},\tau)-\zeta_{\downarrow}({\bf{r}},\tau)\right)$. Thus the second term in Eq.(\ref{Equation_15}) gives a nonlinear density difference term.
Then, after combining with the chemical potential term of the $\zeta$-sublattice, we can write the real space Hubbard-Stratanovich transformation
\begin{eqnarray}
&&e^{-U/4\sum_{{\bf{r}}}\int^{\beta}_{0}d\tau\left(n_{\zeta}({\bf{r}},\tau)-\frac{2\mu'_{\zeta}}{U}\right)^{2}}\sim
\nonumber\\
&&\sim \int\left[DV_{\zeta}\right]e^{\sum_{i}\int^{\beta}_{0}d\tau\left[-\left(\frac{V_{\zeta}({\bf{r}},\tau)}{\sqrt{U}}\right)^{2}+iV_{\zeta}({\bf{r}},\tau)\left(n_{\zeta}({\bf{r}},\tau)-\frac{2\mu'_{\zeta}}{U} \right)\right]},
\nonumber\\
\label{Equation_16}
\end{eqnarray}
where $\mu'_{\zeta}$ is the shifted chemical potential due to the intralayer interaction $U$ and $\mu'_{\zeta}=\mu_{\zeta}+U/2$.  The integration variables $V_{\zeta}({\bf{r}},\tau)$ are the auxiliary field variables introduced at each lattice site ${\bf{r}}$ and time $\tau$, coupled to the electron density term $n_{\zeta}({{\bf{r}},\tau})$ (for example, for the layer 1, $n_{\zeta}({{\bf{r}},\tau})=\sum_{\sigma}a^{\dag}_{\sigma}({\bf{r}},\tau)a_{\sigma}({\bf{r}},\tau)$ for $\zeta=a$, or $\sum_{\sigma}b^{\dag}_{\sigma}({\bf{r}},\tau)b_{\sigma}({\bf{r}},\tau)$, for $\zeta=b$).
The field integral, in the right-hand side (r.h.s.), in Eq.(\ref{Equation_16}), can be evaluated by the steepest descent method. We get
\begin{eqnarray}
\int\left[DV_{\zeta}\right]e^{\sum_{{\bf{r}}}\int^{\beta}_{0}d\tau\left[-\left(\frac{V_{\zeta}({\bf{r}},\tau)}{\sqrt{U}}\right)^{2}+iV_{\zeta}({\bf{r}},\tau)\left(n_{\zeta}({\bf{r}},\tau)-\frac{2\mu_{1}}{U} \right)\right]}\sim
\nonumber\\
\sim e^{-U/2\sum_{{\bf{r}}}\int^{\beta}_{0}d\tau \left(\bar{n}_{\zeta}-\frac{2\mu_{\zeta}}{U}\right)\left(n_{\zeta}({\bf{r}},\tau)-\frac{2\mu_{\zeta}}{U} \right)}.
\nonumber\\
\label{Equation_17}
\end{eqnarray}
Here, in order to obtain the r.h.s. in Eq.(\ref{Equation_17}), we have replaced the field integration over $V_{\zeta}({\bf{r}},\tau)$ by the value of the function in the exponential at the saddle-point of the decoupling potential, i.e., by $\upsilon_{\zeta}=iU/2\left(\bar{n}_{\zeta}-2\mu_{\zeta}/U\right)$, where the average density $\bar{n}_{\zeta}$ is defined as $\bar{n}_{\zeta}=\left\langle n_{\zeta,\uparrow}({\bf{r}},\tau)+n_{\zeta,\downarrow}({\bf{r}},\tau)\right\rangle$. The averages here are defined with the help of the partition function of the  system, given in Eq.(\ref{Equation_11}), i.e.,
\begin{eqnarray}
\left\langle ...\right\rangle=\frac{1}{Z}\int\left[D\bar{X}DX\right]\left[D\bar{Y}DY\right]...e^{-S\left[\bar{X},X,\bar{Y},Y\right]}.
\label{Equation_18}
\end{eqnarray}
The described above  procedure should be repeated also for all nonlinear density terms in the twisted BLG structure, containing the non-linearities in $n_{a}({\bf{r}})$, $n_{b}({\bf{r}})$ $n_{\tilde{a}}({\bf{r}})$,  and $n_{\tilde{b}}({\bf{r}})$. The decoupling of the nonlinear density-difference term, appearing after the transformation, given in Eq.(\ref{Equation_15}), is also straightforward. Namely, for the $\zeta$-type sublattice variables we have
\begin{widetext}
	\begin{eqnarray}
	e^{U\sum_{{\bf{r}}}\int^{\beta}_{0}d\tau \left(S_{\zeta,z}({\bf{r}},\tau)\right)^{2}}=e^{U/4\sum_{{\bf{r}}}\int^{\beta}_{0}d\tau \left(n_{\zeta,\uparrow}({\bf{r}},\tau)-n_{\zeta,\downarrow}({\bf{r}},\tau)\right)^{2}}\sim
	\nonumber\\
	\sim \int{\left[D\Delta_{c,\zeta}\right]}e^{\sum_{{\bf{r}}}\int^{\beta}_{0}d\tau \left[-\left(\frac{\Delta_{c,\zeta}({\bf{r}},\tau)}{\sqrt{U}}\right)^{2}+\Delta_{c,\zeta}({\bf{r}},\tau)\left(n_{\zeta,\uparrow}({\bf{r}},\tau)-n_{\zeta,\downarrow}({\bf{r}},\tau)\right)\right]}.
	\label{Equation_19}
	\end{eqnarray}
\end{widetext}
Next, after functional differentiation, the saddle-point value of the variables $\Delta_{c,\zeta}({\bf{r}},\tau)$ is given as $\delta_{c,\zeta}=U/2\left\langle n_{\zeta,\uparrow}({\bf{r}},\tau)-n_{\zeta,\downarrow}({\bf{r}},\tau)\right\rangle$. Thus, it is proportional to the difference between the electron densities with the opposite spin polarizations. For simplicity, we suppose the case of the spin balanced BLG layers, with equal density numbers for each spin direction, i.e. $\left\langle n_{\zeta,\uparrow}({\bf{r}},\tau)\right\rangle=\left\langle n_{\zeta,\downarrow}({\bf{r}},\tau)\right\rangle$, and the quantities $\delta_{c,\zeta}$ vanish in this case: $\delta_{c,\zeta}=0$. 
%
\subsection{\label{sec:Section_2_2} The Hubbard interlayer interaction term $W_{\theta}$}
%
It is not difficult to show that the density product part in the last interaction term in the Hamiltonian, given in Eq.(\ref{Equation_8}) could be rewritten as
\begin{eqnarray}
H_{W}=-W_{\theta}\sum_{{\bf{r}},\sigma\sigma'}\left[|{\chi}_{a\tilde{a}}({\bf{r}},\tau,\sigma;{\bf{r}}+\delta{\bf{r}}+d{\bf{e}}_{z},\tau,\sigma')|^{2}\right.
\nonumber\\
\left.+|{\chi}_{b\tilde{b}}({\bf{r}},\tau,\sigma;{\bf{r}}+\delta{\bf{r}}+d{\bf{e}}_{z},\tau,\sigma')|^{2}\right],
\label{Equation_20}
\end{eqnarray}
where we have introduced the new complex variables ${\chi}_{a\tilde{a}}({\bf{r}},\tau,\sigma;{\bf{r}}+\delta{\bf{r}}+d{\bf{e}}_{z},\tau,\sigma')$, ${\chi}_{b\tilde{b}}({\bf{r}},\tau,\sigma;{\bf{r}}+\delta{\bf{r}}+d{\bf{e}}_{z},\tau,\sigma')$ for different sublattices and their complex conjugates ${\chi}^{\dag}_{a\tilde{a}}({\bf{r}},\tau,\sigma;{\bf{r}}+\delta{\bf{r}}+d{\bf{e}}_{z},\tau,\sigma')$, ${\chi}^{\dag}_{b\tilde{b}}({\bf{r}},\tau,\sigma;{\bf{r}}+\delta{\bf{r}}+d{\bf{e}}_{z},\tau,\sigma')$. These new operators create the electron-hole pairs between different layers in the twisted bilayer graphene. They are defined as  
\begin{eqnarray}
{\chi}_{a\tilde{a}}({\bf{r}},\tau,\sigma;{\bf{r}}+\delta{\bf{r}}+d{\bf{e}}_{z},\tau,\sigma')=
\nonumber\\
=\tilde{a}^{\dag}_{\sigma'}({\bf{r}}+\delta{\bf{r}}+d{\bf{e}}_{z},\tau)a_{\sigma}({\bf{r}},\tau),
\nonumber\\
{\chi}_{b\tilde{b}}({\bf{r}},\tau,\sigma;{\bf{r}}+\delta{\bf{r}}+d{\bf{e}}_{z},\tau,\sigma')=
\nonumber\\
=\tilde{b}^{\dag}_{\sigma'}({\bf{r}}+\delta{\bf{r}}+d{\bf{e}}_{z},\tau)b_{\sigma}({\bf{r}},\tau).
\nonumber\\
\label{Equation_21}
\end{eqnarray}
The interlayer interaction part of the interaction Hamiltonian $H_{U-W}$, given in the form, in Eq.(\ref{Equation_20}), is more for convenient for further decoupling procedure. We give here the real-space linearization for the term which couples the sublattice sites $A$, and $\tilde{A}$, in tBLG. This is given by the first term in Eq.(\ref{Equation_20}). Here, we apply the complex form of the Hubbard-Stratanovich transformation \cite{cite_43} for the one-component fermion-field by introducing the new external source fields ${\Xi}^{\dag}_{a\tilde{a}}({\bf{r}},\tau,\sigma;{\bf{r}}+\delta{\bf{r}}+d{\bf{e}}_{z},\tau,\sigma')$ and ${{\Xi}}_{a\tilde{a}}({\bf{r}},\tau,\sigma;{\bf{r}}+\delta{\bf{r}}+d{\bf{e}}_{z},\tau,\sigma')$, coupled to the operators ${\chi}_{a\tilde{a}}({\bf{r}},\tau,\sigma;{\bf{r}}+\delta{\bf{r}}+d{\bf{e}}_{z},\tau,\sigma')$ and ${\chi}^{\dag}_{a\tilde{a}}({\bf{r}},\tau,\sigma;{\bf{r}}+\delta{\bf{r}}+d{\bf{e}}_{z},\tau,\sigma')$, respectively, i.e., we have
\begin{widetext}
  \begin{eqnarray}
&&e^{W_{\theta}\sum_{{\bf{r}},\sigma,\sigma'}\int^{\beta}_{0}d\tau|{\chi}_{a\tilde{a}}({\bf{r}},\tau,\sigma;{\bf{r}}+\delta{\bf{r}}+d{\bf{e}}_{z},\tau,\sigma')|^{2}}=
\int{\left[D\bar{\Xi}_{a\tilde{a}}D\Xi_{a\tilde{a}}\right]}\exp\left[\sum_{{\bf{r}},\sigma,\sigma'}\int^{\beta}_{0}d\tau \left(-\frac{|{\Xi}_{a\tilde{a}}({\bf{r}},\tau,\sigma;{\bf{r}}+\delta{\bf{r}}+d{\bf{e}}_{z},\tau,\sigma')|^{2}}{W_{\theta}}+\right.\right.
\nonumber\\
&&\left.\left.+\bar{\Xi}_{a\tilde{a}}({\bf{r}},\tau,\sigma;{\bf{r}}+\delta{\bf{r}}+d{\bf{e}}_{z},\tau,\sigma')\chi_{a\tilde{a}}({\bf{r}},\tau,\sigma;{\bf{r}}+\delta{\bf{r}}+d{\bf{e}}_{z},\tau,\sigma')\right.\right.
\nonumber\\
&&\left.\left.+\bar{\chi}_{a\tilde{a}}({\bf{r}},\tau,\sigma;{\bf{r}}+\delta{\bf{r}}+d{\bf{e}}_{z},\tau,\sigma'){\Xi}_{a\tilde{a}}({\bf{r}},\tau,\sigma;{\bf{r}}+\delta{\bf{r}}+d{\bf{e}}_{z},\tau,\sigma')\frac{}{}\right)\right].
\nonumber\\
\label{Equation_22}
  \end{eqnarray}
  \end{widetext}
The same linearization procedure could be done also for the second term in the sum, in Eq.(\ref{Equation_20}).  
Then after taking the functional derivative of the partition function with respect to the source field ${\Xi}^{\dag}_{a\tilde{a}}({\bf{r}},\tau,\sigma;{\bf{r}}+\delta{\bf{r}}+d{\bf{e}}_{z},\tau,\sigma')$, i.e.,
 \begin{eqnarray}
\frac{\delta{Z}}{\delta{{\Xi}^{\dag}_{a\tilde{a}}}}=0,
 \label{Equation_23}
   \end{eqnarray}
we will get the saddle point value of the decoupling field variables ${\Xi}_{a\tilde{a}}({\bf{r}},\tau,\sigma;{\bf{r}}+\delta{\bf{r}}+d{\bf{e}}_{z},\tau,\sigma')$ which gives us exactly the excitonic pairing gap parameter $\Delta^{a\tilde{a}}_{\sigma\sigma'}$ in tBLG
\begin{eqnarray}
\Delta^{a\tilde{a}}_{\sigma\sigma'}={\Xi}^{\rm s.p.}_{a\tilde{a}}=W_{\theta}\left\langle\tilde{a}^{\dag}_{\sigma'}({\bf{r}}+\delta{\bf{r}}+d{\bf{e}}_{z},\tau)a_{\sigma}({\bf{r}},\tau)\right\rangle.
\label{Equation_24}
\end{eqnarray}
In the same way, we get the excitonic gap parameter $\Delta^{b\tilde{b}}_{\sigma\sigma'}$ for the electron and hole quasiparticles at the sublattice sites $\tilde{B}$ and $B$, in different layers. We have
\begin{eqnarray}
\Delta^{b\tilde{b}}_{\sigma\sigma'}={\Xi}^{\rm s.p.}_{b\tilde{b}}=W_{\theta}\left\langle\tilde{b}^{\dag}_{\sigma'}({\bf{r}}+\delta{\bf{r}}+d{\bf{e}}_{z},\tau)b_{\sigma}({\bf{r}},\tau)\right\rangle.
\label{Equation_25}
\end{eqnarray}
Supposing the isotropic structure of the bilayer graphene with the non-deformed crystall structure, we have $\Delta^{a\tilde{a}}_{\sigma\sigma'}=\Delta^{b\tilde{b}}_{\sigma\sigma'}\equiv \Delta_{\sigma\sigma'}$. Furthermore, we consider here the homogeneous tBLG structure with the pairing between the particles with the same orientation of spin variables, i.e. $\Delta_{\sigma\sigma'}=\Delta_{\sigma}\delta_{\sigma\sigma'}$. 
%
\subsection{\label{sec:Section_2_3} Fermionic action in ${\bf{k}}$-space}
%
Furthermore, we can write the total action of the fermion system, in the Fourier-space representation, given by the transformations $\eta_{\ell,\sigma}({\bf{r}},\tau)=\frac{1}{\beta{N}}\sum_{{\bf{k}},\nu_{n}}\eta_{\sigma}({\bf{k}},\nu_{n})e^{i\left({\bf{k}}{\bf{r}}-\nu_{n}\tau\right)}$, where $\nu_{n}=\pi\left(2n+1\right)/\beta$, with $n=0,\pm1,\pm2,...$, are the fermionic Matsubara frequencies \cite{cite_44}, and $N$ is the total number of sites in the $\eta$-type sublattice, in the layer $\ell$. 

Next, we introduce the four component Nambu-spinors for the rotated bilayer graphene at each discrete state ${\bf{k}}$, in the reciprocal space, and for a given spin direction $\sigma=\uparrow$ or $\sigma=\downarrow$ 
\begin{eqnarray} 
\footnotesize
\arraycolsep=0pt
\medmuskip = 0mu
{\psi}_{\sigma}({\bf{k}},\nu_{n})=\left(
\begin{array}{crrrr}
a_{{\bf{k}}+\frac{{\bf{\Delta{k}}}}{2},\sigma}(\nu_{n})\\\\
b_{{\bf{k}}+\frac{{\bf{\Delta{k}}}}{2},\sigma}(\nu_{n}) \\\\
\tilde{a}_{{\bf{k}}-\frac{{\bf{\Delta{k}}}}{2},\sigma}(\nu_{n}) \\\\
\tilde{b}_{{\bf{k}}-\frac{{\bf{\Delta{k}}}}{2},\sigma}(\nu_{n})
\end{array}
\right),
\label{Equation_26}
   \end{eqnarray}
where ${\Delta{{\bf{k}}}}$ is the rotation vector in ${\bf{k}}$-space ${\Delta{{\bf{k}}}}=2\sin(\theta/2)\left({{\bf{k}}}\times{\bf{e}}_{z}\right)$. Then the total action of the system, given in Eq.(\ref{Equation_12}) will be rewritten in the following form 
 \begin{widetext}
 \begin{eqnarray}
&&S\left[a,b,\tilde{a},\tilde{b}\right]=-\frac{1}{\beta{N}}\sum_{{\bf{k}},\nu_n}\sum_{\sigma}\left({\gamma}_{{\bf{k}}+\frac{\Delta{\bf{k}}}{2}}{a}^{\dag}_{{\bf{k}}+\frac{\Delta{\bf{k}}}{2},\sigma}(\nu_n){b}_{{\bf{k}}+\frac{\Delta{\bf{k}}}{2},\sigma}(\nu_n)+h.c.\right)
\nonumber\\
&&-\frac{1}{\beta{N}}\sum_{{\bf{k}},\nu_n}\sum_{\sigma}\left(\tilde{{\gamma}}_{{\bf{k}}-\frac{\Delta{\bf{k}}}{2}}\tilde{a}^{\dag}_{{\bf{k}}-\frac{\Delta{\bf{k}}}{2},\sigma}(\nu_n)\tilde{b}_{{\bf{k}}-\frac{\Delta{\bf{k}}}{2},\sigma}(\nu_n)+h.c.\right)-\frac{\tilde{\gamma}_1}{\beta{N}}\sum_{{\bf{k}},\nu_n}\sum_{\sigma}\left(a^{\dag}_{{\bf{k}}+\frac{\Delta{\bf{k}}}{2},\sigma}(\nu_n)\tilde{a}_{{\bf{k}}-\frac{\Delta{\bf{k}}}{2},\sigma}(\nu_n)+h.c.\right)
\nonumber\\
&&-\frac{\tilde{\gamma}_1}{\beta{N}}\sum_{{\bf{k}},\nu_n}\sum_{\sigma}\left(b^{\dag}_{{\bf{k}}+\frac{\Delta{\bf{k}}}{2},\sigma}(\nu_n)\tilde{b}_{{\bf{k}}-\frac{\Delta{\bf{k}}}{2},\sigma}(\nu_n)+h.c.\right)-\frac{\mu_{\rm 1eff}}{\beta{N}}\sum_{{\bf{k}},\nu_n}\sum_{\sigma}\left(n_{a\sigma}\left({\bf{k}}+\frac{\Delta{\bf{k}}}{2},\nu_n\right)+n_{b\sigma}\left({\bf{k}}+\frac{\Delta{\bf{k}}}{2},\nu_n\right)\right)
\nonumber\\
&&-\frac{\mu_{\rm 2eff}}{\beta{N}}\sum_{{\bf{k}},\nu_n}\sum_{\sigma}\left(n_{\tilde{a}\sigma}\left({\bf{k}}-\frac{\Delta{\bf{k}}}{2},\nu_n\right)+n_{\tilde{b}\sigma}\left({\bf{k}}-\frac{\Delta{\bf{k}}}{2},\nu_n\right)\right)
-\frac{1}{\beta{N}}\sum_{{\bf{k}},\nu_{n}}\sum_{\sigma}\left(\Delta^{\dag}_{\sigma}\tilde{a}^{\dag}_{{\bf{k}}-\frac{\Delta{\bf{k}}}{2},\sigma}(\nu_n)a_{{\bf{k}}+\frac{\Delta{\bf{k}}}{2},\sigma}(\nu_n)+h.c.\right)
\nonumber\\
&&-\frac{1}{\beta{N}}\sum_{{\bf{k}},\nu_{n}}\sum_{\sigma}\left(\Delta^{\dag}_{\sigma}\tilde{b}^{\dag}_{{\bf{k}}-\frac{\Delta{\bf{k}}}{2},\sigma}(\nu_n)b_{{\bf{k}}+\frac{\Delta{\bf{k}}}{2},\sigma}(\nu_n)+h.c.\right).
\label{Equation_27}
\end{eqnarray}
\end{widetext}
The effective chemical potentials $\mu_{\rm 1eff}$ and $\mu_{\rm 2eff}$, appearing in Eq.(\ref{Equation_27}) are defined as $\mu_{\rm 1eff}=\mu+U/4$ and $\mu_{\rm 2eff}=\mu+U/4+W_{\theta}$. We have supposed here the equal values of the chemical potentials in both layers of the initially untwisted bilayer graphene, i.e., $\mu_{1}=\mu_{2}\equiv\mu$. The ${\bf{k}}$-dependent parameters ${\gamma}_{{\bf{k}}+\frac{\Delta{\bf{k}}}{2}}$ and $\tilde{{\gamma}}_{{\bf{k}}-\frac{\Delta{\bf{k}}}{2}}$ are the energy dispersion parameters for the lower and top layer, respectively, in the non-interacting tBLG. They are defined as follows
\begin{eqnarray}
{\gamma}_{{\bf{k}}+\frac{\Delta{\bf{k}}}{2}}=&&\gamma_{0}\left[\frac{}{}e^{-i(k_{x}+\sin(\frac{\theta}{2})k_{y})a}+2e^{i(k_{x}+\sin(\frac{\theta}{2})k_{y})\frac{a}{2}}\times\right.
\nonumber\\
&&\left.\times\cos\frac{a\sqrt{3}}{2}(k_{y}-k_{x}\sin({\theta}/{2}))\right]
\nonumber\\
\label{Equation_28}
\end{eqnarray}
and the energy dispersion parameter for the top layer $\ell=2$ in the twisted bilayer graphene is given as
\begin{eqnarray}
&&\tilde{\gamma}_{{\bf{k}}-\frac{\Delta{\bf{k}}}{2}}=\gamma_0\left(e^{i({\bf{k}}-{\Delta{\bf{k}}}/{2})\bm{\mathit{\delta}}'_1}+e^{i({\bf{k}}-{\Delta{\bf{k}}}/{2})\bm{\mathit{\delta}}'_2}+\right.
\nonumber\\
&&\left.e^{i({\bf{k}}-{\Delta{\bf{k}}}/{2})\bm{\mathit{\delta}}'_3}\right)=\gamma_0\left[e^{-i(k_{x}f_{1}(\theta)+k_{y}f_{2}(\theta))a}+\right.
\nonumber\\
&&\left.+2e^{i(k_{x}f_{1}(\theta)+k_{y}f_{2}(\theta))\frac{a}{2}}\cos\frac{a\sqrt{3}}{2}(k_{x}f_{2}(\theta)-k_{y}f_{1}(\theta))\right].
\nonumber\\
\label{Equation_29}
\end{eqnarray}
The functions $f_{1}(\theta)$ and $f_{2}(\theta)$, introduced in Eq.(\ref{Equation_29}), were defined as
 \begin{eqnarray}
f_{1}(\theta)=\cos(\theta)+\sin(\theta)\sin(\frac{\theta}{2}),
\nonumber\\
f_{2}(\theta)=\sin(\theta)-\cos(\theta)\sin(\frac{\theta}{2}).
\label{Equation_30}
\end{eqnarray}
Then, with the help of the Nambu spinors, introduced in Eq.(\ref{Equation_26}), we can rewrite the total fermionic action of the tBLG system in the following form
\begin{eqnarray} 
 S\left[{\psi}^{\dag},\psi,\theta\right]=\frac{1}{\beta{N}}\sum_{{\bf{k}},\nu_{n}}\sum_{\sigma}{\psi}^{\dag}_{\sigma}({\bf{k}},\nu_{n}){\cal{G}}^{-1}_{\sigma}({\bf{k}},\nu_{n}){\psi}_{\sigma}({\bf{k}},\nu_{n}),
\nonumber\\
  \label{Equation_31}
   \end{eqnarray}
where, ${\cal{G}}^{-1}_{\sigma}({\bf{k}},\nu_{n})$, is the inverse Green's function matrix, of size $4\times4$. It is defined as
\begin{eqnarray}
&&{\cal{G}}^{-1}_{\sigma}({\bf{k}},\nu_{n})=
\nonumber\\
&&=-\left(
\begin{array}{ccccrrrr}
E_{1}(\nu_{n}) & \gamma_{{\bf{k}}+\frac{\Delta{\bf{k}}}{2}} & \tilde{\gamma}_1+\Delta_{\sigma} & 0\\
\gamma^{\dag}_{{\bf{k}}+\frac{\Delta{\bf{k}}}{2}} &E_{1}(\nu_{n})  & 0 &  \tilde{\gamma}_1+\Delta_{\sigma}\\
\tilde{\gamma}_1+\Delta^{\dag}_{\sigma} & 0 & E_{2}(\nu_{n}) & \tilde{\gamma}_{{\bf{k}}-\frac{\Delta{\bf{k}}}{2}} \\
0 & \tilde{\gamma}_1+\Delta^{\dag}_{\sigma} & \tilde{\gamma}^{\dag}_{{\bf{k}}-\frac{\Delta{\bf{k}}}{2}}  & E_{2}(\nu_{n}) 
\end{array}
\right).
\nonumber\\
\label{Equation_32}
\end{eqnarray}
The diagonal elements in the matrix, in Eq.(\ref{Equation_32}), are the energy parameters $E_{1}(\nu_{n})=i\nu_{n}+\mu_{1\rm eff}$ and $E_{2}(\nu_{n})=i\nu_{n}+\mu_{2\rm eff}$ and the effective chemical potentials $\mu_{1\rm eff}$ and $\mu_{2\rm eff}$, are defined with the help of the intralayer and interlayer interaction parameters $U$ and $W_{\theta}$, as it has been discussed previousely, in this Section. We assume here that the pairing gap is real and is not spin-dependent ($\Delta_\sigma\equiv\Delta=\bar{\Delta}$). Therefore, the structure of the Green's function matrix does not changes for the opposite spin direction: $\hat{G}^{-1}_{\downarrow}\left({\bf{k}},\nu_{n}\right)\equiv \hat{G}^{-1}_{\uparrow}\left({\bf{k}},\nu_{n}\right)$. The partition function of the tBLG system will be rewritten as
\begin{eqnarray}
Z=\int\left[{\cal{D}}{\psi}^{\dag}{\cal{D}}\psi\right]e^{-S\left[{\psi}^{\dag},\psi,\theta\right]}.
\label{Equation_33}
\end{eqnarray}
Furthermore, we introduce the external source fields and we obtain the generating functional by applying the Hubbard-Stratanovich transformation in Eq.(\ref{Equation_33}), i.e., integrating over Nambu-fields. We have
\begin{widetext}
\begin{eqnarray}
G\left[J^{\dag},J\right]=&&\int\left[{\cal{D}}{\psi}^{\dag}{\cal{D}}\psi\right]e^{-\frac{1}{2}\sum_{{\bf{k}},\nu_n,\sigma}\psi^{\dag}_{\sigma}({\bf{k}},\nu_n,\sigma)\tilde{{\cal{G}}}^{-1}_{\sigma}({\bf{k}},\nu_n)\psi_{\sigma}({\bf{k}},\nu_n)}
e^{\frac{1}{2}\sum_{{\bf{k}},\nu_n}\sum_{\sigma}J^{\dag}_{\sigma}({\bf{k}},\nu_n)\psi_{\sigma}({\bf{k}},\nu_n)+\frac{1}{2}\sum_{{\bf{k}},\nu_n}\sum_{\sigma}\psi^{\dag}_{\sigma}({\bf{k}},\nu_n)J_{\sigma}({\bf{k}},\nu_n)}
\nonumber\\
&&\sim \exp{\frac{1}{2}\sum_{{\bf{k}},\nu_n}\sum_{\sigma}J^{\dag}_{\sigma}({\bf{k}},\nu_n){\cal{G}}_{\sigma}({\bf{k}},\nu_n)J_{\sigma}({\bf{k}},\nu_n)},
\label{Equation_34}
\end{eqnarray}
\end{widetext}
where $J_{\sigma}({\bf{k}},\nu_n)$ is the four dimensional source field vector defined for the subattices $A$, $B$ , $\tilde{A}$ and $\tilde{B}$, analogous to the Nambu vectors in Eq.(\ref{Equation_26}). Next, $\tilde{{\cal{G}}}^{-1}_{\sigma}({\bf{k}},\nu_n)$ is the inverse Green's function matrix introduced as $\tilde{{\cal{G}}}^{-1}_{\sigma}({\bf{k}},\nu_n)=(2/\beta{N}){\cal{G}}^{-1}_{\sigma}({\bf{k}},\nu_n)$ with ${\cal{G}}^{-1}_{\sigma}({\bf{k}},\nu_n)$, given in Eq.(\ref{Equation_32}), above. The direct Green's function matrix, in the right hand side in Eq.(\ref{Equation_34}), is obtained by inverting the matrix form ${\cal{G}}^{-1}_{\sigma}({\bf{k}},\nu_n)$, i.e., ${\cal{G}}_{\sigma}({\bf{k}},\nu_n)=\left(\tilde{{\cal{G}}}^{-1}_{\sigma}({\bf{k}},\nu_n)\right)^{-1}$. By differentiating the generating functional $G\left[J^{\dag},J\right]$ with respect to source field variables $J^{\dag}$ and $J$ we can calculate different grand canonical averages, describing the physical properties of the system tBLG, such as the partial average electron densities, chemical potential and excitonic pairging gap parameter (see in the next Section). 
%
\subsection{\label{sec:Section_2_4} The chemical potential}
%
The half-filling condition in the tBLG system gives the following equation for the chemical potential 
\begin{eqnarray}
2\left(\left\langle n_{a\uparrow}({\bf{r}},\tau)\right\rangle+\left\langle n_{\tilde{a}\uparrow}({\bf{r}}+\delta{{\bf{r}}}+d{\bf{e}}_{z},\tau)\right\rangle\right)-1=0.
\label{Equation_35}
\end{eqnarray}
Here, we have taken into account the fact that
\begin{eqnarray}
\left\langle n_{a\uparrow}\right\rangle=\left\langle n_{b\uparrow}\right\rangle,
\nonumber\\
\left\langle n_{\tilde{a}\uparrow}\right\rangle=\left\langle n_{\tilde{b}\uparrow}\right\rangle.
\label{Equation_36}
\end{eqnarray}
Then, after Fourier transformation, we can write for the partial average electron densities at the sublattice sites $A$ and $\tilde{A}$ 
\begin{eqnarray}
\left\langle n_{a}({\bf{r}},\tau)\right\rangle=\frac{1}{(\beta{N})^{2}}\sum_{{\bf{k}},\nu_{n}}\sum_{\sigma}\left\langle a^{\dag}_{\sigma}({\bf{k}},\nu_{n})a_{\sigma}({\bf{k}},\nu_{n})\right\rangle,
\nonumber\\
\left\langle n_{\tilde{a}}({\bf{r}},\tau)\right\rangle=\frac{1}{(\beta{N})^{2}}\sum_{{\bf{k}},\nu_{n}}\sum_{\sigma}\left\langle {\tilde{a}}^{\dag}_{\sigma}({\bf{k}},\nu_{n})\tilde{a}_{\sigma}({\bf{k}},\nu_{n})\right\rangle.
\label{Equation_37}
\end{eqnarray}
Then we use the expression of the generating functional in Eq.(\ref{Equation_34}) to calculate explicitely the averages in Eq.(\ref{Equation_37})
\begin{eqnarray}
\left\langle a^{\dag}_{\sigma}({\bf{k}},\nu_{n})a_{\sigma}({\bf{k}},\nu_{n})\right\rangle=2{\cal{G}}_{11}({\bf{k}},\nu_{n})
\nonumber\\
=\beta{N}\frac{{\cal{A}}_{11}({\bf{k}},\nu_n)}{\det{{\cal{G}}}^{-1}({\bf{k}},\nu_{n})},
\nonumber\\
\left\langle {\tilde{a}}^{\dag}({\bf{k}},\nu_{n}){\tilde{a}}({\bf{k}},\nu_{n})\right\rangle=2{\cal{G}}_{33}({\bf{k}},\nu_{n})
\nonumber\\
=\beta{N}\frac{{\cal{A}}_{33}({\bf{k}},\nu_n)}{\det{{\cal{G}}}^{-1}({\bf{k}},\nu_{n})},
\label{Equation_38}
\end{eqnarray}
where the functions ${\cal{A}}_{11}({\bf{k}},\nu_n)$ and ${\cal{A}}_{33}({\bf{k}},\nu_n)$ in nominators, in r.h.s., are defined as 
\begin{eqnarray}
{\cal{A}}_{11}({\bf{k}},\nu_n)=-\left[(i\nu_n)^{3}+g_1(i\nu_n)^{2}+g_2(i\nu_n)+g_3\right]
\label{Equation_39}
\end{eqnarray}
with the coefficients $g_1$, $g_2$, and $g_{3}$, in Eq.(\ref{Equation_39}), defined as
\begin{eqnarray}
g_1=\mu_{\rm 1eff}+2\mu_{\rm 2eff},
\nonumber\\
g_2=2\mu_{\rm 1eff}\mu_{\rm 2eff}+\mu^{2}_{\rm 2eff}-(\Delta+\tilde{\gamma_1})^{2}-|\tilde{\gamma}_{{\bf{k}}-\frac{\Delta{{\bf{k}}}}{2}}|^{2},
\nonumber\\
g_3=\mu_{\rm 1eff}\mu^{2}_{\rm 2eff}-\mu_{\rm 2eff}(\Delta+\tilde{\gamma_1})^{2}-\mu_{\rm 1eff}|\tilde{\gamma}_{{\bf{k}}-\frac{\Delta{{\bf{k}}}}{2}}|^{2}
\label{Equation_40}
\end{eqnarray}
and
\begin{eqnarray}
{\cal{A}}_{33}({\bf{k}},\nu_n)=-\left[(i\nu_n)^{3}+g_4(i\nu_n)^{2}+g_5(i\nu_n)+g_6\right]
\label{Equation_41}
\end{eqnarray}
with the coefficients $g_4$, $g_5$ and $g_6$ defined as
\begin{eqnarray}
g_4=2\mu_{\rm 1eff}+\mu_{\rm 2eff},
\nonumber\\
g_5=2\mu_{\rm 1eff}\mu_{\rm 2eff}+\mu^{2}_{\rm 1eff}-(\Delta+\tilde{\gamma_1})^{2}-|{\gamma}_{{\bf{k}}+\frac{\Delta{{\bf{k}}}}{2}}|^{2},
\nonumber\\
g_6=\mu_{\rm 2eff}\mu^{2}_{\rm 1eff}-\mu_{\rm 1eff}(\Delta+\tilde{\gamma_1})^{2}-\mu_{\rm 2eff}|{\gamma}_{{\bf{k}}+\frac{\Delta{{\bf{k}}}}{2}}|^{2}.
\label{Equation_42}
\end{eqnarray}
It is clear that the partial averages $\left\langle n_{a}({\bf{r}},\tau)\right\rangle$ and $\left\langle n_{\tilde{a}}({\bf{r}},\tau)\right\rangle$, being defined as the normal fermionic Green's functions, have the dimension of the inverse of energy $\dim\left[n_{\eta}\right]=1/\varepsilon$, 
Then we get for the sum of the partial average densities in Eq.(\ref{Equation_37}) the following expression
\begin{eqnarray}
\left\langle n_{a} \right\rangle+\left\langle n_{b} \right\rangle=-\frac{2}{\beta{N}}\sum_{{\bf{k}},\nu_{n}}\sum^{4}_{i=1}\frac{\alpha_{i{\bf{k}}}}{i\nu_{n}-\varepsilon_{i}({\bf{k}})},
\label{Equation_43}
\end{eqnarray}
where the dimensionless coefficients $\alpha_{i{\bf{k}}}$ are given as
\begin{eqnarray}
\footnotesize
\arraycolsep=0pt
\medmuskip = 0mu
\alpha_{i{{\bf{k}}}}=(-1)^{i+1}
\left\{
\begin{array}{cc}
\displaystyle  & \frac{{\cal{P}} ^{(3)}(\varepsilon_{i{\bf{k}}})}{\varepsilon_{1{\bf{k}}}-\varepsilon_{2{\bf{k}}}}\prod_{j=3,4}\frac{1}{\varepsilon_{i{\bf{k}}}-\varepsilon_{j{\bf{k}}}},  \ \ \  $if$ \ \ \ i=1,2,
\newline\\
\newline\\
\displaystyle  & \frac{{\cal{P}} ^{(3)}(\varepsilon_{3{\bf{k}}})}{\varepsilon_{3{\bf{k}}}-\varepsilon_{4{\bf{k}}}}\prod_{j=1,2}\frac{1}{\varepsilon_{i{\bf{k}}}-\varepsilon_{j{\bf{k}}}} ,  \ \ \  $if$ \ \ \ i=3,4,
\end{array}\right.
\nonumber\\
\label{Equation_44}
\end{eqnarray}
Here, ${\cal{P}}^{(3)}(\varepsilon_{i{\bf{k}}})$ is the third order polynomial in $\varepsilon_{i{\bf{k}}}$ defined as

\begin{eqnarray}
{\cal{P}} ^{(3)}(\varepsilon_{i{\bf{k}}})=2\varepsilon^{3}_{i{\bf{k}}}+g_{7}\varepsilon^{2}_{i{\bf{k}}}+g_{8}\varepsilon_{i{\bf{k}}}+g_{9}.
\label{Equation_45}
\end{eqnarray}
The coefficients $g_{7}$, $g_8$ and $g_9$ are given with the help of the coefficients in Eqs.(\ref{Equation_40}) and (\ref{Equation_42}). Particularly, we have
\begin{eqnarray}
&&g_{7}=g_{1}+g_{4}=3(\mu_{\rm 1eff}+\mu_{\rm 2eff}),
\nonumber\\
&&g_{8}=g_{2}+g_{5}=(\mu_{\rm 1eff}+\mu_{\rm 2eff})^{2}+2\left[\mu_{\rm 1eff}\mu_{\rm 2eff}\right.
\nonumber\\
&&
\left.-\left(\Delta+\tilde{\gamma}_{1}\right)^{2}\right]
-|\gamma_{{\bf{k}}+\frac{\Delta{{\bf{k}}}}{2}}|^{2}-|\tilde{\gamma}_{{\bf{k}}-\frac{\Delta{{\bf{k}}}}{2}}|^{2},
\nonumber\\
&&g_{9}=g_{3}+g_{6}=\left(\mu_{\rm 1eff}\mu_{\rm 2eff}-\left[\Delta+\tilde{\gamma}_{1}\right)^{2}\right]\times
\nonumber\\
&&\times\left(\mu_{\rm 1eff}+\mu_{\rm 2eff}\right)-\mu_{\rm 1eff}|\tilde{\gamma}_{{\bf{k}}-\frac{\Delta{{\bf{k}}}}{2}}|^{2}-\mu_{\rm 2eff}|\gamma_{{\bf{k}}+\frac{\Delta{{\bf{k}}}}{2}}|^{2}.
\nonumber\\
\label{Equation_46}
\end{eqnarray}
The energy parameters $\varepsilon_{i{\bf{k}}}$ define the interacting band structure in twisted bilayer graphene with the excitonic pairing mechanism. They are the solutions of the fourth order algebraic equation 
\begin{eqnarray}
\det\left[\tilde{{\cal{G}}}^{-1}_{\sigma}({\bf{k}},\nu_{n})\right]=0. 
 \label{Equation_47}
 \end{eqnarray}
Furthermore, we put the expression in Eq.(\ref{Equation_43}) into Eq.(\ref{Equation_35}) and we get the equation for the chemical potential $\mu$
 \begin{eqnarray}
 -\frac{2}{N}\sum_{{\bf{k}},\nu_{n}}\sum^{4}_{i=1}\alpha_{i{\bf{k}}}n_{F}\left(\varepsilon_{i{\bf{k}}}+\mu\right)=1.
 \label{Equation_48}
 \end{eqnarray}
The function $n_{F}(x)$ is the Fermi-Dirac distribution function $n_{F}(x)=1/\left(e^{\beta(x-\mu)}+1\right)$ and the summation over $i$ is due to the four solutions of the equation for the determinant of the inverse of the Green's function matrix, given in Eq.(\ref{Equation_47}).
%
 \subsection{\label{sec:Section_2_5} The equation for the gap}
 %
 The excitonic gap parameter is determined in Eqs.(\ref{Equation_24}) and (\ref{Equation_25}) for the pairing between the particles at the $A$-$\tilde{A}$ and $B$-$\tilde{B}$ sublattice sites positions. After the Fourier transformation into the reciprocal space, those equations take the following forms
\begin{eqnarray}
&&\Delta^{a\tilde{a}}=
\nonumber\\
&&=\frac{W_{\theta}}{({\beta{N}})^{2}}\sum_{{\bf{k}},\nu_{n}}\sum_{\sigma}\left\langle\tilde{a}^{\dag}_{{\bf{k}}-\frac{\Delta{{\bf{k}}}}{2},\sigma}\left(\nu_{n}\right)a_{{\bf{k}}+\frac{\Delta{{\bf{k}}}}{2},\sigma}\left(\nu_{n}\right)\right\rangle,
\nonumber\\
&&\Delta^{b\tilde{b}}=
\nonumber\\
&&=\frac{W_{\theta}}{({\beta{N}})^{2}}\sum_{{\bf{k}},\nu_{n}}\sum_{\sigma}\left\langle\tilde{b}^{\dag}_{{\bf{k}}-\frac{\Delta{{\bf{k}}}}{2},\sigma}\left(\nu_{n}\right)b_{{\bf{k}}+\frac{\Delta{{\bf{k}}}}{2},\sigma}\left(\nu_{n}\right)\right\rangle.
\label{Equation_49}
\end{eqnarray}
As we have indicated in the previous section, those gap parameters are equal, i.e., $\Delta^{a\tilde{a}}_{\sigma}=\Delta^{b\tilde{b}}_{\sigma}$, for the homogeneous system considered here. We present here the calculation of the gap parameter $\Delta^{a\tilde{a}}_{\sigma}$. The calculation of the order parameter $\Delta^{b\tilde{b}}_{\sigma}$ is very similar and does not represents any difficulty. Again, by using the expression of the generating functional in Eq.(\ref{Equation_34}), we get
\begin{eqnarray}
&&\left\langle\tilde{a}^{\dag}_{{\bf{k}}-\frac{\Delta{{\bf{k}}}}{2},\sigma}\left(\nu_{n}\right)a_{{\bf{k}}+\frac{\Delta{{\bf{k}}}}{2},\sigma}\left(\nu_{n}\right)\right\rangle
=2{\cal{G}}_{13}({\bf{k}},\nu_{n})=
\nonumber\\
&&=\beta{N}\frac{{\cal{A}}_{13}({\bf{k}},\nu_n)}{\det{{\cal{G}}}^{-1}({\bf{k}},\nu_{n})}.
\label{Equation_50}
\end{eqnarray}
Here, the coefficient ${\cal{A}}_{13}({\bf{k}},\nu_n)$ is obtained by inverting the $4\times 4$ Green's function matrix, given in Eq.(\ref{Equation_32}). We have
\begin{eqnarray}
{\cal{A}}_{13}({\bf{k}},\nu_n)=\left(\Delta+\tilde{\gamma}_1\right)\left[(i\nu_{n})^{2}+(i\nu_{n})(\mu_{\rm 1eff}+\mu_{\rm 2eff})\right.
\nonumber\\
\left.+\gamma_{{\bf{k}}+\frac{\Delta{\bf{k}}}{2}}\tilde{\gamma}^{\ast}_{{\bf{k}}-\frac{\Delta{\bf{k}}}{2}}-\left(\Delta+\tilde{\gamma}_1\right)^{2}+\mu_{\rm 1eff}\mu_{\rm 2eff}\right].
\label{Equation_51}
\end{eqnarray}
Similarly, we can write the equation for the complex conjugate parameter $\bar{\Delta}^{a\tilde{a}}_{\sigma}$. We have
\begin{eqnarray}
\bar{\Delta}^{a\tilde{a}}=\frac{W_{\theta}}{(\beta{N})^{2}}\sum_{{\bf{k}},\nu_{n}}\sum_{\sigma}\left\langle{a}^{\dag}_{{\bf{k}}+\frac{\Delta{{\bf{k}}}}{2},\sigma}\left(\nu_{n}\right)\tilde{a}_{{\bf{k}}-\frac{\Delta{{\bf{k}}}}{2},\sigma}\left(\nu_{n}\right)\right\rangle.
\nonumber\\
\label{Equation_52}
\end{eqnarray}
For the average under the sum in Eq.(\ref{Equation_52}) we get 
\begin{eqnarray}
\left\langle{a}^{\dag}_{{\bf{k}}+\frac{\Delta{{\bf{k}}}}{2},\sigma}\left(\nu_{n}\right)\tilde{a}_{{\bf{k}}-\frac{\Delta{{\bf{k}}}}{2},\sigma}\left(\nu_{n}\right)\right\rangle
=2{\cal{G}}_{31}({\bf{k}},\nu_{n})
\nonumber\\
=\beta{N}\frac{{\cal{A}}_{31}({\bf{k}},\nu_n)}{\det{{\cal{G}}}^{-1}({\bf{k}},\nu_{n})}
\label{Equation_53}
\end{eqnarray}
and for the coefficient ${\cal{A}}_{31}({\bf{k}},\nu_n)$, we have 
\begin{eqnarray}
&&{\cal{A}}_{31}({\bf{k}},\nu_n)=\left(\Delta+\tilde{\gamma}_1\right)\left[(i\nu_{n})^{2}+(i\nu_{n})(\mu_{\rm 1eff}+\mu_{\rm 2eff})+\right.
\nonumber\\
&&\left.\gamma^{\ast}_{{\bf{k}}+\frac{\Delta{\bf{k}}}{2}}\tilde{\gamma}_{{\bf{k}}-\frac{\Delta{\bf{k}}}{2}}-\left(\Delta+\tilde{\gamma}_1\right)^{2}+\mu_{\rm 1eff}\mu_{\rm 2eff}\right].
\label{Equation_54}
\end{eqnarray}
From the computational point of view it is much more easier to calculate the sum $\Delta_{\sigma}+\Delta^{\dag}_{\sigma}$, bacause of the mixed products of types $\gamma_{{\bf{k}}}\tilde{\gamma}^{\ast}_{{\bf{k}}}$ and $\gamma^{\ast}_{{\bf{k}}}\tilde{\gamma}_{{\bf{k}}}$ in Eqs.(\ref{Equation_51}) and $(\ref{Equation_54})$. Supposing the real value of the excitonic gap parameter $\Delta_{\sigma}$, we can write
\begin{eqnarray}
\Delta+\Delta^{\dag}=2\Delta=\frac{W_{\theta}\left(\Delta+\tilde{\gamma}_{1}\right)}{{\beta{N}}}\times
\nonumber\\
\times\sum_{{\bf{k}},\nu_{n}}\frac{{\cal{A}}_{13}({\bf{k}},\nu_n)+{\cal{A}}_{31}({\bf{k}},\nu_n)}{\det\left[{\cal{G}}^{-1}({\bf{k}}.\nu_{n})\right]},
\label{Equation_55}
\end{eqnarray}
we have omitted the spin indices in Eq.(\ref{Equation_54}) because of the spin symmetry of the action, given in Eq.(\ref{Equation_31}) and the Green's function matrix in Eq.(\ref{Equation_32}).
For the sum ${\cal{A}}_{13}({\bf{k}},\nu_n)+{\cal{A}}_{31}({\bf{k}},\nu_n)$ we obtain 
\begin{eqnarray}
{\cal{A}}_{13}({\bf{k}},\nu_n)+{\cal{A}}_{31}({\bf{k}},\nu_n)=2(i\nu_{n})^{2}+g_{10}(i\nu_{n})+g_{11},
\label{Equation_56}
\end{eqnarray}
where we have introduced the new coefficients $g_{10}$ and $g_{11}$ as follows
\begin{eqnarray}
&&g_{10}=2\left(\mu_{\rm 1eff}+\mu_{\rm 2eff}\right),
\nonumber\\
&&g_{11}=2\mu_{\rm 1eff}\mu_{\rm 2eff}-2\left(\Delta+\tilde{\gamma}_{1}\right)^{2}
\nonumber\\
&&+\gamma^{\ast}_{{\bf{k}}+\frac{\Delta{\bf{k}}}{2}}\tilde{\gamma}_{{\bf{k}}-\frac{\Delta{\bf{k}}}{2}}+\gamma_{{\bf{k}}+\frac{\Delta{\bf{k}}}{2}}\tilde{\gamma}^{\ast}_{{\bf{k}}-\frac{\Delta{\bf{k}}}{2}}.
\label{Equation_57}
\end{eqnarray}
Again, after performing the Matsubara summation over $\nu_{n}$, we obtain the self-consistent equation for the excitonic gap parameter in twisted bilayer graphene
\begin{eqnarray}
\Delta=\frac{W_{\theta}\left(\Delta+\tilde{\gamma}_{1}\right)}{N}\sum_{{\bf{k}}}\sum^{4}_{i=1}\beta_{i{\bf{k}}}n_{F}\left(\varepsilon_{i{\bf{k}}}+\mu\right),
\label{Equation_58}
\end{eqnarray}
where the coefficients $\beta_{i{\bf{k}}}$ are given by
\begin{eqnarray}
\footnotesize
\arraycolsep=0pt
\medmuskip = 0mu
\beta_{i{{\bf{k}}}}=(-1)^{i+1}
\left\{
\begin{array}{cc}
\displaystyle  & \frac{{\cal{P}}^{(2)}(\varepsilon_{i{\bf{k}}})}{\varepsilon_{1{\bf{k}}}-\varepsilon_{2{\bf{k}}}}\prod_{j=3,4}\frac{1}{\varepsilon_{i{\bf{k}}}-\varepsilon_{j{\bf{k}}}},  \ \ \  $if$ \ \ \ i=1,2,
\newline\\
\newline\\
\displaystyle  & \frac{{\cal{P}}^{(2)}(\varepsilon_{3{\bf{k}}})}{\varepsilon_{3{\bf{k}}}-\varepsilon_{4{\bf{k}}}}\prod_{j=1,2}\frac{1}{\varepsilon_{i{\bf{k}}}-\varepsilon_{j{\bf{k}}}} ,  \ \ \  $if$ \ \ \ i=3,4,
\end{array}\right.
\nonumber\\
\label{Equation_59}
\end{eqnarray}
and ${\cal{P}}^{(2)}(\varepsilon_{i{\bf{k}}})$ is the second order polynomial defined as
\begin{eqnarray}
{\cal{P}} ^{(2)}(\varepsilon_{i{\bf{k}}})=2\varepsilon^{2}_{i{\bf{k}}}+g_{10}\varepsilon_{i{\bf{k}}}+g_{11}.
\label{Equation_60}
\end{eqnarray}
In order to obtain the chemical potential and the excitonic gap parameter, we should solve simultaneously the self-consistent equations in Eqs. (\ref{Equation_48}) and (\ref{Equation_58}).  
%
\section{\label{sec:Section_3} Numerical results}
\begin{figure}
	\begin{center}
		\includegraphics[scale=0.65]{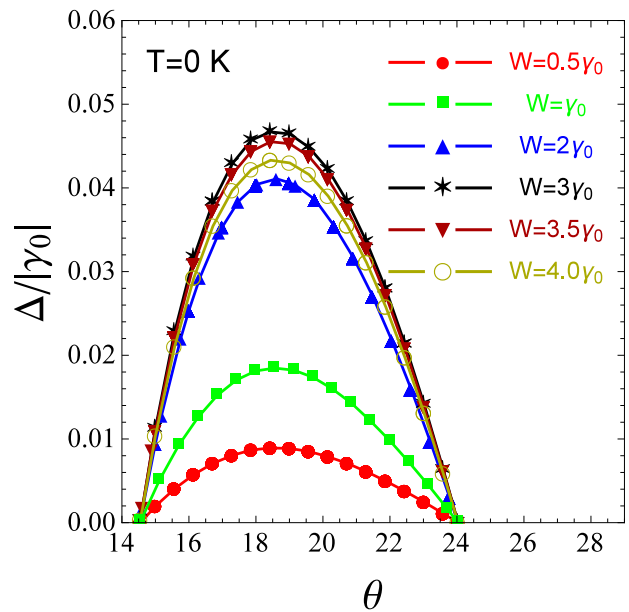}
		\caption{\label{fig:Fig_2}(Color online) The excitonic gap as a function of the twist angle $\theta$ between the layers in the bilayer graphene. The gap parameter is normalized to the intralayer hopping amplitude $\gamma_0$ and the temperature is set at $T=0$. Different values of the local interlayer Coulomb interaction parameter are considered, corresponding to the initially untwisted $AA$-stacked bilayer graphene structure.}
	\end{center}
\end{figure}

%
In this section, we give the numerical results for the excitonic gap parameter and the chemical potential. In Fig.~\ref{fig:Fig_2}, we have presented the solution for the excitonic gap parameter $\Delta$, normalized to the intralayer hopping integral $\gamma_0$. Particularly, the excitonic gap parameter is shown as a function of the twist angle $\theta$ for different values of the interlayer Coulomb interaction parameter $W$. We remember here that the value of the interlayer Coulomb interaction parameter is the local value of it, corresponding to the untwisted bilayer graphene structure, as it has been supposed to be initially. Furthermore, this local value of the interaction gets modified according to the formula given in Eq.(\ref{Equation_9}), where the instantaneous value of the interlayer interaction parameter $W_{\theta}$ is evaluated as a function of the local interaction parameter $W$ and the twist angle $\theta$. The plots in Fig.~\ref{fig:Fig_2} are done for $T=0$ and the intralayer Coulomb interaction parameter $U$ is chosen to be of the order of the hopping amplitude $\gamma_0$, i.e., $U=\gamma_0$. The full strength of the local parameter $W$ has been considered in the plots. We see that the excitonic order parameter $\Delta$ exists only for a certain interval of twist angle $\theta$, mainly, it is clear from the results in Fig.~\ref{fig:Fig_2} that $\Delta\neq0$ for $\theta\in(14^{\circ}, 24^{\circ})$. The excitonic insulator states appear for each twist angle in this interval and the excitonic insulator state corresponds to the regions under the closed paths, given in Fig.~\ref{fig:Fig_2}. We observe a dynamical evolution of the excitonic gap parameter for different values of the interaction parameter $W$. Particularly, for the values of $W$ from the interval $W\in(0.5\gamma_0,3\gamma_0)$ the order parameter $\Delta$ increases gradually, while for the higher values of $W$ it decreases. Next, in this section, we will discuss also the influence of twisting on the electronic band structure in the bilayer graphene. 
%
%
\begin{figure}
	\begin{center}
		\includegraphics[scale=0.65]{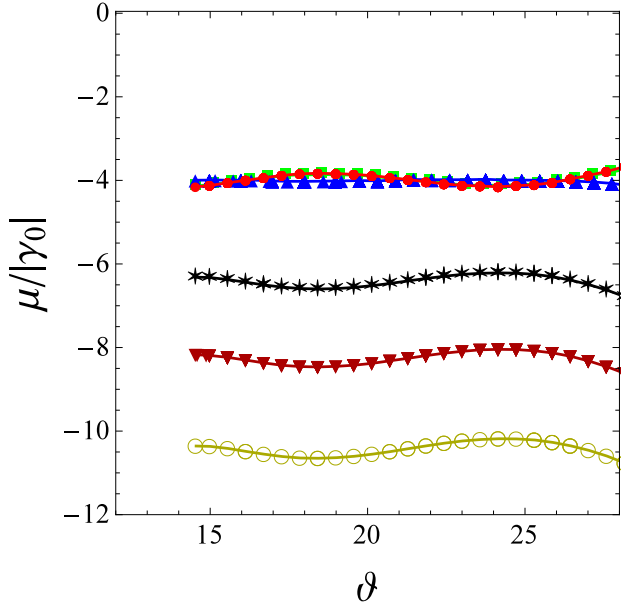}
		\caption{\label{fig:Fig_3}(Color online) The chemical potential in the twisted bilayer graphene as a function of the twist angle $\theta$ between the layers in the bilayer graphene. The chemical potential is normalized to the intralayer hopping amplitude $\gamma_0$ and the temperature is set at $T=0$. Different values (from weak to very strong) of the local interlayer Coulomb interaction parameter are considered, corresponding to the initially untwisted $AA$-stacked bilayer graphene structure.}
	\end{center}
\end{figure} 
%
In Fig.~\ref{fig:Fig_3}, we presented the exact solution of the chemical potential in tBLG calculated self-consistently in Eqs.(\ref{Equation_48}) and Eq.(\ref{Equation_58}). The same interval of the twist angles is considered as in the case of the order parameter and for the same values of the local interlayer interaction parameter $W$. We see that for the values $W=0.5\gamma_0$, $\gamma_0$ and $W=2\gamma_0$ the solutions for the chemical potential $\mu$ are practically coinciding, while for the values $W=3\gamma_0$, $W=3.5\gamma_0$ and $W=4\gamma_0$ the solutions of $\mu$ are strongly distinct. This behavior is exactly the opposite to the behavior of the excitonic order parameter $\Delta$, given in Fig.~\ref{fig:Fig_2}, where the solutions of the order parameter are strongly distinct for small values of the interaction parameter $W$. The main reason for such a counter behavior for $\mu$ and $\Delta$ is that the small variations of the chemical potential lead to the substantial changes in the dynamics of the excitonic order parameter and, the opposite, large changes of $\mu$, as a function of $W$, lead to the small changes in the gap parameter behavior (for a given twist angle $\theta$). This effect is absolutely absent in the usual Bernall-stacked bilayer graphene structure, as it has been discussed in details, in Refs.\onlinecite{cite_41,cite_42}. Another important observation that could be gained from Fig.~\ref{fig:Fig_2} is that the maximum values of the excitonic order parameter $\Delta_{\rm max}$ correspond to the interval $\theta\in(18^{\circ},19^{\circ})$ and this is true for different values of the interlayer interaction parameter $W$. 
%
\begin{figure}
	\begin{center}
		\includegraphics[scale=0.65]{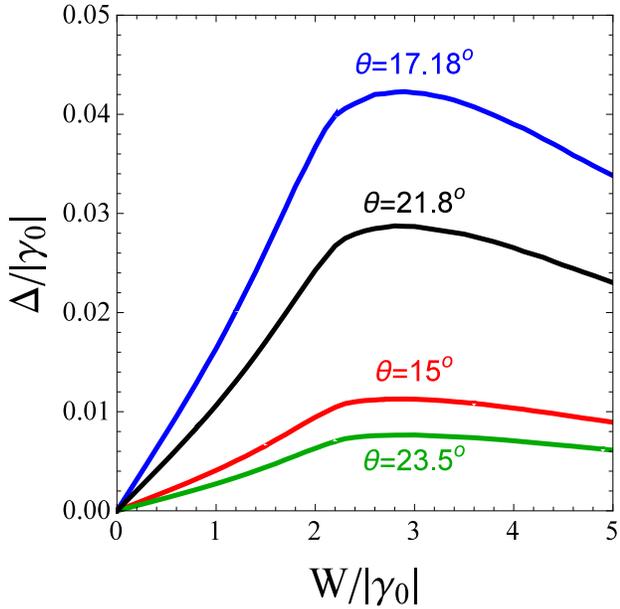}
		\caption{\label{fig:Fig_4}(Color online) The excitonic gap parameter in the twisted bilayer graphene as a function of the interlayer Coulomb interaction parameter $W$ and for different values of the twist angle $\theta$. The excitonic gap parameter is normalized to the intralayer hopping amplitude $\gamma_0$, and the zero temperature case is considered in the picture.}
	\end{center}
\end{figure} 
%
\begin{figure}
	\begin{center}
		\includegraphics[scale=0.65]{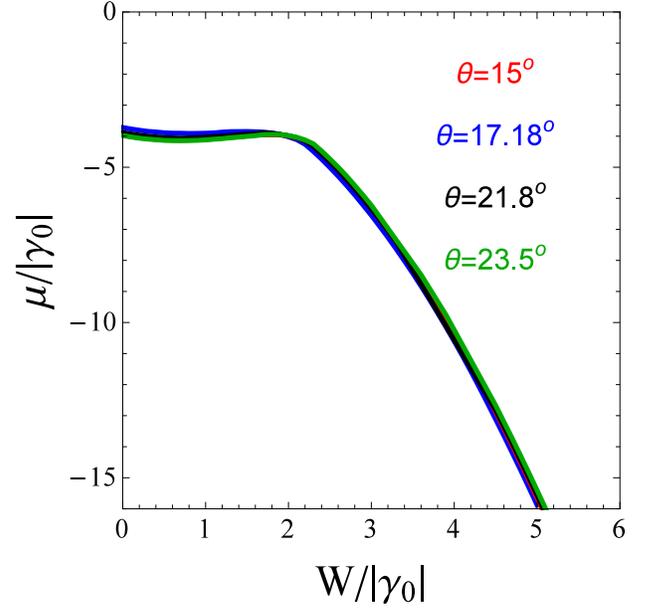}
		\caption{\label{fig:Fig_5}(Color online) The chemical potential in the twisted bilayer graphene as a function of the interlayer Coulomb interaction parameter $W$ and for different values of the twist angle $\theta$. The chemical potential is normalized to the intralayer hopping amplitude $\gamma_0$, and the zero temperature case is considered in the picture.}
	\end{center}
\end{figure} 
%
In Fig.~\ref{fig:Fig_4}, we showed the $W$-dependence of the gap parameter $\Delta$ for different fixed values of the twist angle $\theta$. The maximum value of the order parameter is situated at $W_{c}=2.8\gamma_0$ for all values of twist angle $\theta$. Here, we observe again the dynamical behavior of the order parameter as a function of the twist angle, as in the case, in Fig.~\ref{fig:Fig_2} above. The $W$-dependence of $\Delta$, corresponding to the special case of the smallest unit moir{\'e} supercell \cite{cite_15, cite_45, cite_46} at the twist angle $\theta=21.8^{\circ}$,  is also presented in Fig.~\ref{fig:Fig_4}. The solutions for the chemical potential $\mu$ as a function of $W$, for the same selected values of the twist angles, is shown in Fig.~\ref{fig:Fig_5}, below. We see in Fig.~\ref{fig:Fig_5} that all curves of the chemical potential remain practically unchanged (with very small variations from each other) when changing the twist angle. Consequently, the very small variations in the chemical potential, for different angles, lead to the large variation of the excitonic gap parameter and we recognize here the similar property as in the case, discussed in the context in Figs.~\ref{fig:Fig_2} and ~\ref{fig:Fig_3} at the small values of the interaction parameter $W$. It is clear in ~\ref{fig:Fig_4}, that we have another manifestation of the excitonic insulator state in the plane $(W, \Delta)$. Thus, concluding the description on the results for the gap parameter, we can distinct two different scenarios where the excitonic insulator state could be enhanced in twisted bilayer graphene: a) when considering the $\theta$-dependence of $\Delta$ for the fixed values of the interlayer interaction parameter $W$, b) when considering the $W$-dependence of the order parameter $\Delta$, for different selected values of the twist angle $\theta$. For both cases, we observe sufficiently large domains where the excitonic insulator state merges in tBLG.  
%
\subsection{\label{sec:Section_3_1} The total energy and the photon's Raman spectrum broadening}
\begin{figure}
	\begin{center}
		\includegraphics[scale=0.8]{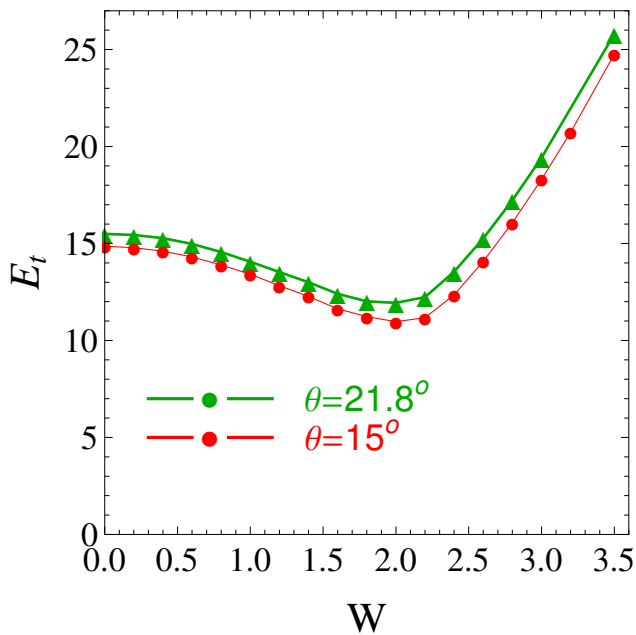}
		\caption{\label{fig:Fig_6}(Color online) The total energy of the twisted bilayer graphene structure as a function of the interlayer Coulomb interaction parameter $W$, for two different values of the twist angle $\theta=21.8^\circ$ and $\theta=15^\circ$. The minimum of the energy corresponds to the formation of the excitonic insulator and excitonic condensate states in tBLG.}
	\end{center}
\end{figure}

%
The total energy of tBLG as a function of the interaction parameter $W$ is given by
\begin{eqnarray}
E_{\rm t}=\frac{1}{N}\sum_{{\bf{k}}}\sum^{4}_{i=1}\varepsilon_{i\bf{k}}.
\label{Equation_61}
\end{eqnarray}
and it is shown in Fig.~\ref{fig:Fig_6} for two different values of the twist angle $\theta$ ($\theta=15^{\circ}$ and $\theta=21.8^{\circ}$). The intralayer Coulomb interaction parameter $U$ fixed at $U=\gamma_0$. Here, for each value of the interaction parameter $W$, we have calculated numerically the excitonic gap parameter $\Delta$ and the chemical potential $\mu$, after which the band structure energies $\varepsilon_{i\bf{k}}$ have calculated as the functionals of those parameters, i.e., $\varepsilon_{i\bf{k}}=\varepsilon_{i\bf{k}}\left[\Delta,\mu,U,W,\theta\right]$ and, furthermore, the sums over ${\bf{k}}$ are effectuated in Eq.(\ref{Equation_61}). 
We see, in Fig.~\ref{fig:Fig_6}, that the total energy $E_{t}$ has deep minima at the value $W=2\gamma_0$. It follows, that this value of the interaction parameter could be more appropriate in order to observe the possible excitonic condensates state in tBLG, in the region of the excitonic insulating phase. Any deviation from the indicated value of $W$ increases the total energy of tBLG, and the corresponding condensate states and excitonic insulator states become metastable states, in the strong matter of fact, although they are possible. In general, the excitonic condensates states are favorable states in tBLG for $W\in(1.5\gamma_0, 2.5\gamma_0)$, and we see also that, at small angle $\theta=15^{\circ}$, the total energy of the tBLG system is lower, for all values of the interaction parameter $W$, by favoring, therefore, the stability of the excitonic condensate. The strong increase of the total energy at $W>2.5\gamma_0$ is closely related to the behavior of the chemical potential in tBLG, presented in Fig.~\ref{fig:Fig_5}, above.  
%
%
\subsection{\label{sec:Section_3_2} Raman spectrum}
\begin{figure}
	\begin{center}
		\includegraphics[scale=0.6]{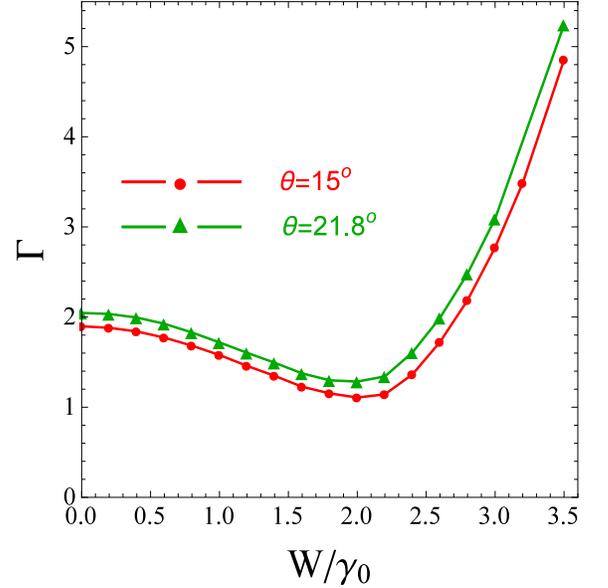}
		\caption{\label{fig:Fig_7}(Color online) Photon-assisted Raman spectrum for the twisted bilayer graphene. Two values of the twist angle have been selected in the picture, corresponding to the total energy curves in Fig.~\ref{fig:Fig_6}. The $W$-dependence of the Raman spectrum is shown in the picture.}
	\end{center}
\end{figure}

%
It has been shown recently \cite{cite_47} that for laser excitations close to the optical transition energies $E_{\rm opt}$, the Raman signal in the tBLG is
enhanced more than an order of magnitude relative to that from BLG with AB stacking. Particularly, a computational approach is reported in Ref.\onlinecite{cite_47}, based on a non-orthogonal tight-binding (NTB) model, for calculation of the electronic density of states, the imaginary part of the dielectric function, and Raman excitation peaks of the ${\bf{G}}$ band in various tBLG. The author calculate the optical transition energies derived from the peaks of the imaginary part of the complex dielectric function. Fitting
a second-order polynomial to the peaks of the dielectric function, the author derives a theoretical dependence of the optical transition energies on the twist angle $\theta$ in tBLG.
Here, we adopt the formula obtained in Ref.\onlinecite{cite_47} for the Raman photon excitation broadening parameter $\Gamma$ and we calculate the $\Gamma$ dependence on $W$ corresponding to the total band structure energies in tBLG obtained in our paper and by considering two different twist angles. Namely, we put (see in Ref.\onlinecite{cite_47}, for details)
\begin{eqnarray}
\Gamma(E)=0.0252E_{\rm t}+0.069E^{2}_{\rm t}.
\label{Equation_62}
\end{eqnarray}
The plots of the photon's Raman broadening parameter $\Gamma$ are shown in Fig.~\ref{fig:Fig_7}. It is obvious that it repeats the behavior of the total energy $E_{\rm t}$ as a function of the parameter $W$, given in Fig.~\ref{fig:Fig_7}. The minima in the spectrum mean that the Raman scattering rate is minimal due to the strong excitonic condensate effects at $W=2\gamma_0$ when the total energy of the system has minimums, and, also, the coefficient $\Gamma$ does not change much with the rotating angle $\theta$. 

\begin{figure}
	\begin{center}
		\includegraphics[scale=0.4]{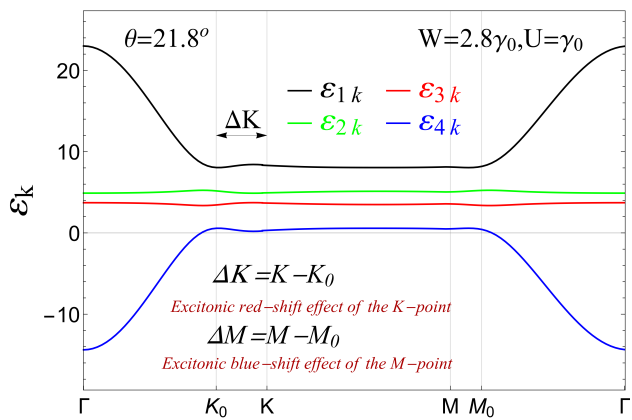}
		\caption{\label{fig:Fig_12}(Color online) Electronic band structure of tBLG at the twist angle $\theta=21.8^\circ$. Different bands are shown with different colors. The interlayer and intralayer Coulomb interaction parameters are chosen as $W=2.8\gamma_0$ and $U=\gamma_0$, respectively. The flat bands $\varepsilon_{2{\bf{k}}}$ and $\varepsilon_{3{\bf{k}}}$ are due to the twist angle between the layers in tBLG, and the appearance of the large band gap is shown in the picture. The excitonic red-shift effect of the principal ${\bf{K}}$-point and the right-shift effect of the $M$-point are shown in the picture.}
	\end{center}
\end{figure}

In Figs.~\ref{fig:Fig_8}, ~\ref{fig:Fig_10}, ~\ref{fig:Fig_11} and ~\ref{fig:Fig_12} we have shown the electronic band structure in tBLG, where the strong excitonic effects and twisting of the layers have been taken into account. In Figs.~\ref{fig:Fig_8} and ~\ref{fig:Fig_10} the electronic band structure is presented for the twisting angle $\theta=15^{\circ}$ and for two different values of the interlayer Coulomb interaction parameter $W$. Particularly, $W=\gamma_0$ (in Fig.~\ref{fig:Fig_8}) and $W=2.8\gamma_0$ (in Fig.~\ref{fig:Fig_10}). It is worth to mention here that we don't supposed the low-${\bf{k}}$ approximation in the theory and the full interaction bandwith has been considered, from low up to very high values of the interaction parameter in twisted BLG.
%
\subsection{\label{sec:Section_3_3} {\bf{K}}-point displacement and doubling}
\begin{figure}
	\begin{center}
		\includegraphics[scale=0.4]{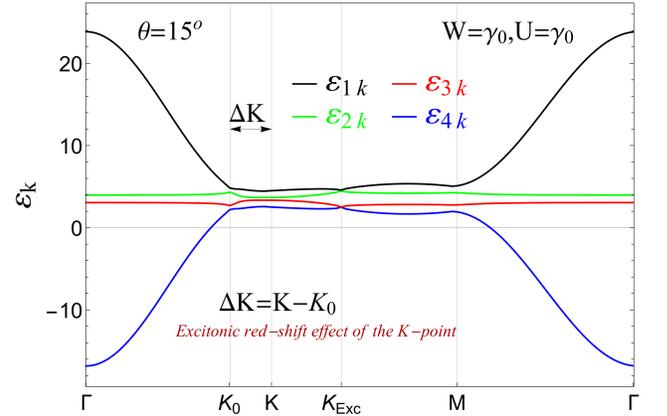}
		\caption{\label{fig:Fig_8}(Color online) Electronic band structure of tBLG at the twist angle $\theta=15^\circ$. Different bands are shown with different colors. The interlayer and intralayer Coulomb interaction parameters are chosen as $W=\gamma_0$ and $U=\gamma_0$, respectively. The flat bands $\varepsilon_{2{\bf{k}}}$ and $\varepsilon_{3{\bf{k}}}$ are due to the twist angle between the layers in tBLG. The excitonic red-shift of the principal ${\bf{K}}$ point and the formation of the additional Dirac's point at $|{\bf{k}}|=K_{\rm Exc}$ are shown in the picture.}
	\end{center}
\end{figure}

%
We have considered the value $W=\gamma_0$, in Fig.~\ref{fig:Fig_8} and the intralayer Coulomb interaction parameter is fixed at the value $U=\gamma_0$. The twist angle is chosen equal to $15^{\circ}$ which corresponds to the small value of the excitonic gap parameter $\Delta$ (see in Figs.~\ref{fig:Fig_2} and ~\ref{fig:Fig_3}). Full four-band electronic structure of the tBLG is shown in the picture and the separate bands are shown by different colors. We see that at the small value of the interlayer Coulomb interaction parameter the system tBLG is in the metallic state due to the twist angle $\theta$. This behavior is exactly opposite to what it has been observed for the AB-Bernal stacked bilayer graphene \cite{cite_42}, where the large band gap is opening at the same value of the Coulomb interaction parameter between the unrotated layers of the BLG. Thus we can conclude here that twisting the BLG leads to the transition in the system from the semiconducting into the metallic state. In addition we see, in Fig.~\ref{fig:Fig_8}, the existence of the nearly flat electronic bands $\varepsilon_{2{\bf{k}}}$ and $\varepsilon_{3{\bf{k}}}$, which are crossing with the bands $\varepsilon_{1{\bf{k}}}$ and $\varepsilon_{4{\bf{k}}}$ respectively, nearly in the middle of the enlarged region $K_{0}\rightarrow M$, namely, at the point $K_{\rm Exc}$ in the first Brillouin zone (FBZ), in the reciprocal space. Such a crossing at $K_{\rm Exc}$ is very unusual in the electronic structure of the tBLG and has no analogs in the existing literature on the subject. This effect could lead furthermore to the very interesting low-frequency spontaneous excitations in the optical excitation spectrum in tBLG. Going further, deep inside in this crossing region, we realize also that the electronic spectrum in the vicinity of the crossing point is linear, thus, leading to an additional Dirac's points on that symmetry region. The observed effect could be explained also in the way that the twisting in the bilayer graphene leads to the doubling of the Dirac point due to the electron-hole correlation effects in the tBLG. We expect that the system tBLG will demonstrate strong metallic behavior along the high symmetry direction $K\rightarrow M$, in the originally defined notations of the BZ of graphene (note here, that the true $K$-point is doubled, and speaking about Dirac's point in the BZ has no anymore the appropriate sense, as it was usually known). Another reason for such an interesting effect could be related to the appearance of the new topological states in the tBLG, mediated by the excitonic gap formation and governed by twisting the BLG layers with the appropriate twist angle. It is interesting to note here that there is a large asymmetry between those doubled $K$-points, $K_{0}$ and $K_{\rm Exc}$, and the true $K$-point position on the ${\bf{k}}$ axis. We attribute such a topological asymmetry with the effect of twisting on the electronic band structure because the excitonic formations could have different binding forces at different excitation levels and positions in the reciprocal space. This question goes far beyond the topics of the presented here theory, namely, here we have to deal with the interfering coherence effects between the electron and hole wave functions, leading to the formation of the excitonic waves which propagate in the reciprocal space. This latest effect is out of the scope of the present paper and needs to be investigated in future \cite{cite_50}.      

%
On the other hand, the flattening effect of the electronic band structure is strongly related to the appearance of the excitonic condensate states in tBLG, which, in turn, is strongly related to the excitonic insulator state, as it has been discussed in details in Ref.\onlinecite{cite_41}. This is very analog to the appearance of the superconducting state in the twisted BLG with the very small magic twist angles, recently discussed in Refs.\onlinecite{cite_36, cite_48, cite_49}, where the twisting leads to the flat bands in the electronic band structure of the tBLG.
%
%

\begin{figure}
	\begin{center}
		\includegraphics[scale=0.8]{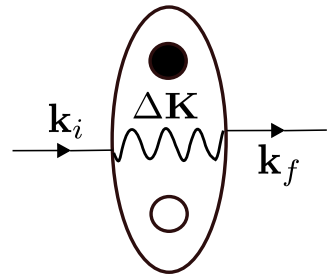}
		\caption{\label{fig:Fig_9}(Color online) Diagram, illustrating the photon's absorption process, in the presence of the excitonic formations.}
	\end{center}
\end{figure}

In Fig.~\ref{fig:Fig_9}, we have presented the diagrammatic process which corresponds to the red-shift effect observed at the principal point $K_{0}$, on the axis $|{\bf{k}}|$ in the BZ. The initial state of the incident photon and its final state are given with the help of the wave vectors ${\bf{k}}_{\rm i}$ and ${\bf{k}}_{\rm f}$, as it is shown in the picture. The excitonic recoil effect is given by the wave vector $\Delta{\bf{K}}$. The quantity  $\Delta{\bf{K}}$ is negative in our case (particularly, $|\Delta{K}|=0.55$ cm$^{-1}$, for the case of low interaction $W=\gamma_0$, discussed in Fig.~\ref{fig:Fig_8}), i.e,  $\Delta{\bf{K}}<0$ which leads to ${\bf{k}}_{\rm f}<{{\bf{k}}_{\rm i}}$, thus we have the Stokes Raman scattering in twisted BLG, when the wavelength of the scattered light is larger than the wavelength of the incident light $\lambda_{\rm f}>\lambda_{\rm i}$. The reason of this is due to the excitonic excitations in the tBLG with the recoil wave vector $\Delta{K}$ which means that the system participates to the emission of a photon with the smaller energy than the original incident wave. The origin of the physics of this effect is very interesting in the context of the Stokes scattering of photons in the tBLG system and should be verified by the appropriate angle-resolved spectroscopy experiments. We should mention here that the value of the Raman scattering rate $\Delta{K}$ is very small, and is slightly larger for the case of the strong interactions $W=2.8\gamma_0$ (see in Fig.~\ref{fig:Fig_10}) at which $|\Delta{K}|=0.67$ cm$^{-1}$. it is also worth to mention that the Stokes scattering rate does not change when varying the twist angle $\theta$.

On the other hand, the anti-Stokes Raman scattering rate takes place at the $M$-point in the BZ, at the high value of the interlayer interaction parameter $W=2.8\gamma_0$, presented in Fig.~\ref{fig:Fig_10}. The blue-shift in the spectrum, associated with the right-shift of the $M$-point, leads to the positive anti-Stokes Raman scattering rate $\Delta{k}>0$ and $\Delta{k}=0.36$ cm$^{-1}$ which is smaller than the Raman's rate associated with the shift of the $K$-point. Here, again, the shift $\Delta{k}$, associated with the $M$-point is independent of $\theta$. Concluding this Section, we remark on the simultaneous existence of the Stokes and anti-Stokes shifts in the electronic band structure in tBLG which will be furthermore verified via the photon Raman scattering measurements.  

The electronic band structure of tBLG, for the higher value of the interlayer Coulomb interaction parameter $W$ and for the same value of the twist angle between the layers in the BLG, is shown in Fig.~\ref{fig:Fig_10}. Namely, we have chosen the value $W=2.8\gamma_0$, which corresponds to the maximum value of the excitonic gap parameter $\Delta_{\rm max}$, for a given value of the twist angle $\theta$ (see in Fig.~\ref{fig:Fig_4}). For example, for the case of $\theta=15^{\circ}$, we have $\Delta_{\rm max}=0.0112708\gamma_0$, and $\mu=-5.6675\gamma_0$. The electronic band structure, in this case, is again flat along the high symmetry direction $K\rightarrow M$, but the difference with the previous case is that now there is a large band gap between the outermost branches in the electronic band structure in tBLG. There are also non-crossing inner bands $\varepsilon_{2{\bf{k}}}$ and $\varepsilon_{3{\bf{k}}}$, and there is sufficiently large band gap also between them. Thus the system is purely in the semiconducting state, and the excitonic condensation is enhanced in the system, being in coexistence with the excitonic insulator state. Thus, tuning only the interlayer Coulomb interaction parameter $W$ could lead to the transition from the metallic into the semiconducting state in the tBLG system (here, we have kept the intralayer Coulomb interaction parameter at the same value, equal to $U=\gamma_0$).    
%
%
Here, again, the position of the true Dirac point $K$, in Fig.~\ref{fig:Fig_10}, is situated deep inside the structure of the flat band, in the FBZ, but there is no doubling effect of the Dirac point. We observe only the remarkable displacement of the Dirac's point region into the left on the ${\bf{k}}$-axis, leading at the same time to the red-shift in the photon's excitation spectrum. Furthermore, the $M$-point is also situated deep inside the flattened region, while the vicinity of the $M$-point is displaced into the right on the ${\bf{k}}$-axis, leading to the left-shift in the excitation spectrum. The displacement of the $M$-point region on the ${\bf{k}}$-axis was absent in the case of the small $W$, discussed above, thus this effect is only governed by $W$, as far as we have observed in Figs.~\ref{fig:Fig_8} and ~\ref{fig:Fig_10}. In this context, it is very important to classify the new $K$-points $K_{0}$ and $K_{\rm Exc}$. First, let's remark that the point $K_{0}$ is a principal stable point, because of its position remains unchanged at the high twist angles which is shown in Fig.~\ref{fig:Fig_11}, where $\theta=21.8^{\circ}$. Contrary, the position of the point $K_{\rm Exc}$ on the ${\bf{k}}$-axis strongly depends on the interaction parameter $W$ and on the twist angle $\theta$, as well. Notably, when changing the twist angle $\theta$ at the fixed value of the interaction parameter $W=\gamma_0$, the excitonic gap parameter $\Delta$ gets changed, which leads, in consequence, to the change in the position of the point $K_{\rm Exc}$ (see in Figs.~\ref{fig:Fig_8} and ~\ref{fig:Fig_11}, in this Section,), i.e., 
\begin{eqnarray}
\Delta{\theta}\rightarrow \delta{\Delta}\rightarrow \delta{K_{\rm exc}}.
\label{Equation_63}
\end{eqnarray}
Thus, the quantity $K_{\rm Exc}$ is a dynamical physical quantity which varies its position depending on the dynamics of the system tBLG, particularly as a function of the angle $\theta$ and the interaction parameter $W$, as it is given in the diagram in Eq.(\ref{Equation_63}).
At the end of this Section, let's remark also that when augmenting the interaction parameter, the whole band structure gets shifted along the band energy axis $\varepsilon_{i{\bf{k}}}$. Particularly, the outermost bands get strongest modifications in that case, while the energy bands $\varepsilon_{2{\bf{k}}}$ and $\varepsilon_{3{\bf{k}}}$ experience small shift effect. The overall displacement in the band structure, which means also that the total system tBLG gets stressed as a unified entity, is related again to the interlayer Coulomb interaction effects. A similar interesting band-shift effect is also found in Ref.\onlinecite{cite_51}, induced by intrinsic electron doping effects. 
%
\subsection{\label{sec:Section_3_4} High twist angles: quartered $K$-point?}
\begin{figure}
	\begin{center}
		\includegraphics[scale=0.4]{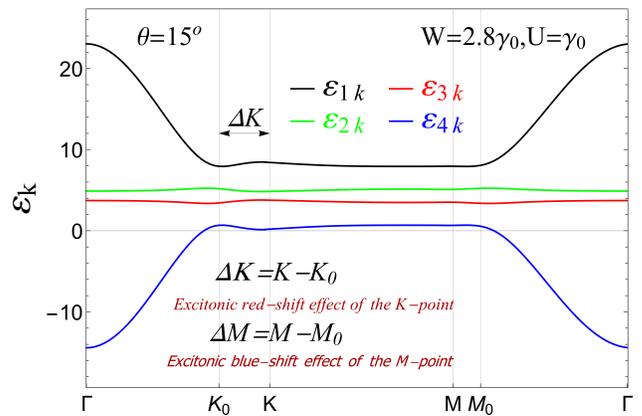}
		\caption{\label{fig:Fig_10}(Color online) Electronic band structure of tBLG at the twist angle $\theta=15^\circ$. Different bands are shown with different colors. The interlayer and intralayer Coulomb interaction parameters are chosen as $W=2.8\gamma_0$ and $U=\gamma_0$, respectively. The flat bands $\varepsilon_{2{\bf{k}}}$ and $\varepsilon_{3{\bf{k}}}$ are due to the twist angle between the layers in tBLG, and the appearance of the large band gap is shown in the picture. The excitonic red-shift effect of the principal ${\bf{K}}$-point and the right-shift effect of the $M$-point are shown in the picture.}
	\end{center}
\end{figure}

\begin{figure}
	\begin{center}
		\includegraphics[scale=0.4]{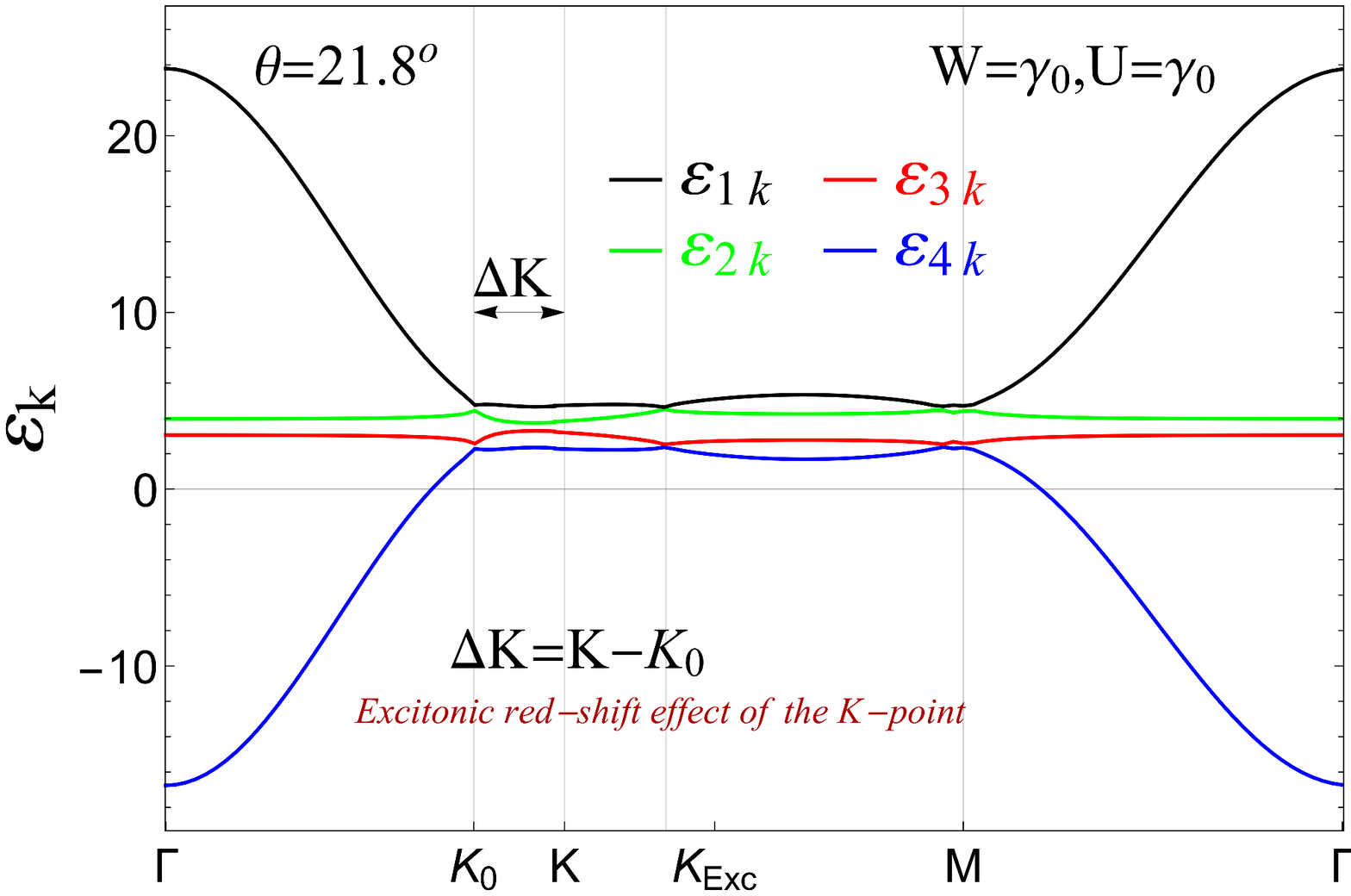}
		\caption{\label{fig:Fig_11}(Color online) Electronic band structure of tBLG at the twist angle $\theta=21.8^\circ$. Different bands are shown with different colors. The interlayer and intralayer Coulomb interaction parameters are chosen as $W=\gamma_0$ and $U=\gamma_0$, respectively. The flat bands $\varepsilon_{2{\bf{k}}}$ and $\varepsilon_{3{\bf{k}}}$ are due to the twist angle between the layers in tBLG, and the appearance of the large band gap is shown in the picture. The excitonic red-shift effect of the principal ${\bf{K}}$-point and the formation of four Dirac's nodes are shown in the figure.}
	\end{center}
\end{figure}

%
A very exciting physics appears when considering the higher value of twist angle between the layers in the BLG and when calculating the electronic band structure with the excitonic pairing effects. Particularly, we have considered the electronic band structure at the twist angle equal to $21.8^{\circ}$ which is the magic angle, corresponding to the smallest moir{\'e} pattern formation unit supercell \cite{cite_15, cite_45, cite_46} in the tBLG. In Fig.~\ref{fig:Fig_10}, we see the main difference in the electronic band structure when augmenting $\theta$, at the small value of the interaction parameter $W=\gamma_0$. The effect of augmenting the twist angle leads, in this case, to the appearance of two additional Dirac's nodes in the vicinity of the $M$-point, apart of doubled Dirac's point along the symmetry direction $K\rightarrow M$. We have called this effect as the quartered $K$-point effect, because, indeed, we get four places at which the Dirac's nodes manifest at the place of the one in the usual band structure picture corresponding to the untwisted and unbiased bilayer graphene. The right-shift of the $K$-point position remains practically the same as in the case of the smaller twist angle $\theta=15^{\circ}$. Meanwhile, the electronic bands do not get displaced as compared with the case of the small rotation angles. The main difference with the case $\theta=15^{\circ}$ is that the position of the point $K_{\rm Exc}$ on the ${|\bf{k}|}$-axis gets shifted to the left on ${|\bf{k}|}$, leading to the additional right-shift in the energy spectrum at the point $K_{\rm Exc}$. Namely, for the case $\theta=15^{\circ}$ we get $K_{\rm Exc}=2.383$ cm$^{-1}$ at $W=\gamma_0$, while for the case $\theta=21.8^{\circ}$ we get $K_{\rm Exc}=2.192$ cm$^{-1}$. This effect is related purely to the change in the twist angle and the interlayer interaction parameter is fixed at the value $W=\gamma_0$. On the other hand, the origin of this shift could be related also with the enhancement of the additional Dirac's nodes at the $M$-point, being the consequence of the propagating excitonic wave function, which has more zeros (nodes) in the case of the high twist angles and small interactions. The doubling or the quartering of the Dirac's point, across the BZ, will undoubtedly lead to the modification of the Fermi surface of the tBLG, however, this is out of the scope of the present discussion. In the high interaction limit and at the high twisting angles, the electronic band structure is nearly exactly the same as in the case of the small interaction limit, and the remarkable effect is that the electronic band structure becomes flattered in this limit.  The calculation result for the case $W=2.8\gamma_0$ (corresponding to the maximum of the excitonic gap parameter $\Delta$) and for the case, $\theta=21.8^{\circ}$ is shown in  Fig.~\ref{fig:Fig_11}. It is remarkable to note here that at the high interaction limit we have only the doubling effect of the $K$-point and there are not Dirac's nodes at the $M$-point in the BZ. Such a property of tBLG could have its applications in the modern nanotechnology and computer science since the bilayer graphene  \cite{cite_41, cite_42} and tBLG will be the ideal candidates for replacing the silicon nanotechnology with the one based on graphene heterostructures. This could revolutionize the quantum microelectronics, optoelectronics, and nanosensing techniques, due to the extremely controllable properties of the twisted bilayer graphene-based heterostructures. The quantum computing domain will also be profited in turn, in this chain.   
%
%
%
%
\section{\label{sec:Section_4} Final Remarks}
%
We have considered the excitonic effects in neutral twisted bilayer graphene structure. The bilayer Hubbard model has been applied for calculating the excitonic gap parameter and chemical potential in tBLG. 
The interlayer hopping amplitude has been modeled according to the Slater-Koster parameterization techniques and the local interlayer Coulomb interaction has been included in the calculations with the help of the full interaction bandwidth parameter. The excitonic gap parameter and the chemical potential have been calculated both analytically and numerically. The excitonic insulator state has been found for a given interval of the twist angle and the excitonic gap parameter has been calculated for different values of the local interlayer Coulomb interaction parameter. The interaction dependence of the excitonic gap parameter is found for different twisting angles. We have included the full interaction bandwidth, starting from the very small up to very high interaction values and no assumptions have been made on the electronic wave vector ${\bf{k}}$ (as the low-${\bf{k}}$ expansions near Dirac's point). For the particular twisting angles, the total energy of the tBLG system has been calculated and the existence of the excitonic condensate state has been predicted in tBLG. The calculations of the electronic band structure, at the small interaction limit, shows the doubling effect of the Dirac's point $K$ at the small values of the rotation angle $\theta$, while at the high values of $\theta$ (namely at $21.8^{\circ}$) we have the quartered Dirac's point and there are four Dirac's nodes in the electronic band structure along the high symmetry directions in the BZ. We have shown the excitonic red-shift effect of the principal $K$-point at the small and high values of the interlayer Coulomb interaction parameter and different twisting angles. At the high values of the interaction parameter, a blue-shift effect of the $M$-point appears in the energy spectrum, which is completely absent in the low interaction limit. We have shown that the existence of the Stokes Raman scattering with the participation of excitons and the mechanism of the photon's absorption in the strongly correlated electron-hole system tBLG, could be explained in connection with the red-shift effect of the principal $K$-point in the electronic band structure.    

The metal-semiconducting transition follows from the dynamical excitonic effects, discussed in the paper. In the low interaction regime, the tBLG system is in the strong metallic regime, without the gap in the excitation spectrum. We have shown that when augmenting the interaction parameter, the system passes into the semiconducting regime with extremely large gap formation in the band structure picture. The observed transition remains true for all values of the twist angle between the layers. Thus, we have shown that the system tBLG is very interesting for the nano- and optoelectronic applications. The rich excitonic physics in tBLG is promising for the wide applications of the tBLG structures in modern solid state physics electronic devices and could revolutionize the related application domains.     
%

%
\end{document}